\documentclass[prc,twocolumn,superscriptaddress,showpacs,psfig,show s]{revtex4}

\usepackage{feynmp}
\usepackage{slashed}
\usepackage{MnSymbol}
\usepackage{amsmath}
\usepackage{amsfonts}
\usepackage{graphicx}
\usepackage{dcolumn}
\usepackage{color}
\usepackage[utf8]{inputenc}
\usepackage{amsthm}
\usepackage{fontenc}
\usepackage{bm}
\RequirePackage{slashed}\usepackage{cancel}
\usepackage[normalem]{ulem}

\usepackage{array,booktabs}
\newcolumntype{C}{>{$\displaystyle}c<{$}}

\begin{document}

\pacs{13.60.Hb,14.20.Dh,27.10.+h}
\keywords{Deeply Virtual Compton scattering, bound proton parton structure}

\title{
Incoherent deeply virtual Compton scattering off $^4$He
}

\author{Sara Fucini}
\affiliation{ Dipartimento di Fisica e Geologia,
Universit\`a degli Studi di Perugia and Istituto Nazionale di Fisica Nucleare,
Sezione di Perugia, via A. Pascoli, I - 06123 Perugia, Italy}
\author{Sergio Scopetta}
\affiliation{ Dipartimento di Fisica e Geologia,
Universit\`a degli Studi di Perugia and Istituto Nazionale di Fisica Nucleare,
Sezione di Perugia, via A. Pascoli, I - 06123 Perugia, Italy}
\author{Michele Viviani}
\affiliation{
INFN-Pisa, 56127 Pisa, Italy} 
\date{\today}

\begin{abstract}
\vspace{0.5cm}
Very recently, for the first time,
the two channels of nuclear deeply virtual Compton scattering, the coherent and incoherent ones, have been separated by the CLAS collaboration 
at the Jefferson Laboratory, using a $^4$He target.
The incoherent channel, which can provide a tomographic view
of the bound proton and shed light on its elusive parton structure, is
thoroughly analyzed here in the Impulse Approximation.
A convolution formula for the relevant 
nuclear cross sections in terms of those
for the bound proton is derived.
Novel scattering amplitudes
for a bound moving nucleon have been obtained and used.
A state-of-the-art
nuclear spectral function, based on the Argonne 18 potential, exact in the two-body part, with the recoiling system in its ground state, and modelled in the remaining contribution,
with the recoiling system in an excited state, has been used. Different parametrizations of the generalized parton distributions of the struck proton have been tested. A good overall agreement with the data for the beam spin asymmetry is obtained. 
It is found that the conventional nuclear effects predicted by the present approach are relevant in 
deeply virtual Compton scattering
and in the competing Bethe-Heitler mechanism,
but they cancel each other to a large extent in their ratio, to which the measured asymmetry is proportional.
Besides, the calculated ratio of the beam spin asymmetry of the bound proton to that of the free one does not describe that estimated by the experimental collaboration. This points to possible interesting effects beyond the Impulse Approximation analysis presented here. 
It is therefore clearly demonstrated that the comparison of the results of a conventional realistic approach, as the one presented here, with future precise data, has the potential
to expose quark and gluon effects in nuclei.
Interesting perspectives for the next measurements at high luminosity facilities, such as JLab at 12 GeV and the future Electron Ion Collider, are addressed.

\end{abstract}

\maketitle
\section{Introduction}
A quantitative understanding of the
European Muon 
collaboration (EMC) effect
in inclusive deep inelastic scattering
(DIS) off nuclear targets
\cite{Aubert:1983xm}
is still missing after
several decades. Since then,
it is clear that
the parton structure of bound nucleons
is modified by the nuclear medium
(see Ref. \cite{Hen:2013oha} for a recent report), but 
so far it has not been possible to distinguish between 
several different explanations,
proposed using different descriptions of the structure of the bound nucleons.
It is widely understood that measurements beyond DIS,
such as semi-inclusive DIS (SIDIS) and nuclear deeply virtual Compton scattering (DVCS), the hard exclusive leptoproduction of a real photon on a nuclear target,
will play a fundamental role in shedding light on this long-standing problem of hadronic Physics
\cite{Dupre:2015jha,Cloet:2019mql}. Crucial steps forward are expected from a new generation of planned measurements at
high energy and high luminosity facilities
in the next years,
including the Jefferson laboratory (JLab) at 12 GeV 
\cite{Dudek:2012vr} and the future electron-ion collider (EIC)
\cite{Accardi:2012qut,Aidala:2020mzt}.
From the theoretical point of view, this
programme implies the challenging description
of complicated processes. One of them,
incoherent DVCS off $^4$He nuclei, for which the first
data have been collected and recently published
\cite{Hattawy:2018liu}, is the subject of this work.

In DVCS, the parton structure is encoded in the so called Compton Form factors (CFFs), defined in terms
of the
generalized parton distributions (GPDs)
\cite{gpds}, non perturbative 
quantities providing a wealth of novel information
(for exhaustive reports see, e.g.,
Ref. \cite{Diehl:2003ny,Belitsky:2005qn,Boffi:2007yc}).
In particular, nuclear DVCS could unveil
the presence of non-nucleonic degrees of freedom
in nuclei \cite{Berger:2001zb}, or
may allow to better understand the spatial distribution of nuclear forces
\cite{Polyakov:2002yz,Polyakov:2018zvc}
(to develope this latter program, the use of positron beams,
presently under discussion at JLab \cite{Accardi:2020swt}, would be of great help). 
Besides,
the tomography of the target, i.e., the distribution
of  partons with a given longitudinal momentum
in the transverse plane, is certainly
one of the most exciting information accessible in DVCS through the GPDs formalism
\cite{Burkardt:2000za}.
In nuclei, DVCS can occur
through two different mechanisms, i.e., the coherent one $A(e,e'\gamma)A$, where the target $A$ recoils
elastically and its tomography can be
ultimately studied, and the incoherent one $A(e,e'\gamma p)X$, where the nucleus breaks up and the struck proton is detected, so that its tomography could be obtained. The comparison between this information and that obtained for the free proton could provide ultimately a pictorial view of the realization of the EMC effect. 
From an experimental point of view, the study of nuclear DVCS requires the very difficult coincidence detection 
of fast photons and electrons
together with slow, intact recoiling protons
or nuclei. 
For this reason, in the first measurement of nuclear DVCS at
HERMES \cite{Airapetian:2009cga}, a clear 
separation between the two different DVCS channels
was not achieved.
Recently, for the first time, such a separation has been performed by the EG6 experiment
of the CLAS collaboration \cite{eric}, with the 6 GeV electron beam at Jefferson Lab (JLab).
The first data for coherent 
and incoherent DVCS off $^4$He have been published in Refs. \cite{Hattawy:2017woc}
and
\cite{Hattawy:2018liu}, respectively. 
Among few nucleon systems, 
for which a realistic evaluation of 
conventional nuclear effects is possible
in principle,
$^4$He is deeply bound and
represents the prototype of a typical finite nucleus. 
Realistic approaches allow to distinguish conventional nuclear effects
from exotic ones, which could be responsible of the observed EMC behaviour. Without realistic benchmark calculations,  
the interpretation of the data will be hardly conclusive. Indeed, in  Refs. \cite{Hattawy:2017woc,Hattawy:2018liu}, the
importance of new calculations has been addressed, for a
successful interpretation of the collected data and of those planned at JLab
in the next years \cite{Armstrong:2017wfw,Armstrong:2017zcm}. 
In facts available estimates, proposed long time ago, correspond
in some cases to different kinematical regions \cite{Guzey:2003jh,Liuti:2005gi}.
New refined calculations are certainly important, above all,
for the next generation of accurate measurements.
In this sense, the use of heavier targets,
due to the difficulty of the corresponding
realistic many-body calculations, is less promising.
Among few-body nuclear systems, $^2$H
is very interesting, for the extraction of the neutron information
and for its rich spin structure
\cite{Berger:2001zb,Cano:2003ju,Taneja:2011sy,Cosyn:2018rdm}.
In between $^2$H and $^4$He, $^3$He could allow
to study the $A$ dependence of nuclear effects and
it could give an easy access to neutron polarization properties,
due to its specific spin structure. Besides, being not isoscalar, 
flavor dependence of nuclear effects could be studied, 
in particular if parallel measurements 
on $^3$H targets were possible.
A complete impulse approximation (IA)
analysis, using the Argonne 18 (AV18)
nucleon-nucleon potential
\cite{Wiringa:1994wb} and the UIX three nucleon force model of Ref. 
\cite{Pudliner:1995wk},
is available and
nuclear effects on GPDs are found to be sensitive to details of 
the used nucleon-nucleon interaction
\cite{Scopetta:2004kj,
Scopetta:2009sn,
Rinaldi:2012pj,Rinaldi:2012ft,
Rinaldi:2014bba}.
Measurements for $^3$He have been addressed, planned in some cases
but they have not been performed yet.
We have therefore analyzed 
successfully, in impulse approximation (IA), 
coherent DVCS off
$^4$He
\cite{Fucini:2018gso}, obtaining an overall good agreement with the data 
\cite{Hattawy:2017woc}.
In a recent rapid communication \cite{prdrc},
we have proposed 
an analogous analysis for the incoherent
channel, to see to what extent
a conventional description can describe
the recent data \cite{Hattawy:2018liu},
which have the tomography
of the bound proton as the ultimate goal.
In that analysis, the incoherent DVCS beam spin asymmetry
has been 
evaluated in IA framework, in terms of
a diagonal spectral function
\cite{Viviani:2001wu}
based 
on the AV18+UIX nuclear interactions
and the GPDs model by Goloskokov and Kroll 
\cite{Goloskokov:2011rd},
obtaining an overall good description of the available data.

We retake here the subject in detail.
The expressions for all the relevant scattering amplitudes for a bound, moving proton are fully derived and explicitly given. In terms of them,
the relevant cross sections are calculated, showing the effects of the use of different descriptions
of the nuclear structure and of the nucleon GPDs. Results are shown for the differential cross sections and the beam spin asymmetry, investigating carefully the source of nuclear effects on both of these observables. 

The paper is structured as follows.
The framework and the main formalism are presented in the next section, while details are collected in two extended appendices.
In the third section, the ingredients of the calculation are described,
while numerical results are presented and discussed in the following one. Conclusions and perspectives are eventually given in the last section.

\section{Formalism}

\begin{figure}[htbp]
\includegraphics[scale=0.40,angle=0]{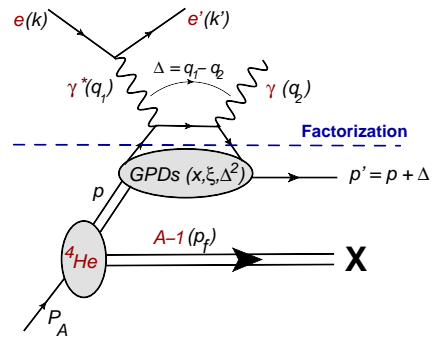}
\caption{(color online)
Incoherent DVCS process off $^4$He in the IA to the handbag approximation.
}
\label{dvcsinco}
\end{figure}

In this section, we present the relevant formalism
for the IA description of
the handbag approximation to 
the incoherent DVCS process  $^4He(e,e' \gamma p')X$, 
shown in Fig. \ref{dvcsinco}.
In such a description of the process, the proton changes its momentum from $p$ to $p'$ after the interaction of the virtual photon with one quark belonging to one nucleon, i.e., only
nucleonic degrees of freedom are included and coherent
effects, such as shadowing, are neglected.
The other IA assumption is that any further scattering between the proton and the remnant system $X$ is disregarded in the final state.
The factorization property can be applied to this process when the initial photon virtuality, $Q^2 = -q_1^2 = -(k-k')^2$, is much larger than the momentum
transferred at hadronic level, {$t = \Delta^2 = (p-p')^2 $}. 
We note also that, in the present IA approach, $  \Delta^2 = (q_1-q_2)^2$, that is, the momentum transferred to the system coincides with that transferred to the struck proton.
For high enough values of $Q^2$, IA usually describes the bulk of nuclear effects in a hard electron scattering process (see,
e.g., Ref. \cite{Slifer:2008re} for an experimental study of the onset of
the validity of IA). Similar expectations hold
in this study, although only the comparison with
data can establish the validity of the chosen framework. 
In this way, the hard vertex of the 
diagram illustrated in Fig. \ref{dvcsinco} can be calculated using perturbative methods while the soft part can be parametrized through the GPDs of the bound proton.
{Such non perturbative objects, namely the GPDs, are functions of $\Delta^2$,}
of the so-called skewness  $\xi =-{\Delta^+}/{P}^+$,
i.e., the difference in plus momentum fraction between the initial and the final states, and of $x$,
the average plus momentum fraction of the struck parton 
with respect to the total momentum.
(the notation $a^\pm = (a_0 \pm a_3)/\sqrt{2}$ is used;
besides, the average four momentum for the photons is $q=(q_1 +q_2)/2$, while
we have defined
$P = p+p'$). 
Actually GPDs, as any other parton dostribution, depend
on the momentum scale $Q^2$ according to QCD evolution equations.
Such an obvious dependence is omitted in the rest of the paper
to avoid a too heavy notation.
We adopted the reference frame proposed in Ref. \cite{Belitsky:2001ns},
with the target at rest,
the virtual photon with energy $\nu$ moving opposite to the 
$\hat{z}$ axis and the leptonic and hadronic planes of the reaction defining the angle $\phi$.
Using energy-momentum conservation, one gets for
the azimuthal angle of the detected proton
the relation $\phi_{p'} = \phi + \phi_e$ and, since in the chosen frame one has, for the electron azimuthal angle, $\phi_e=0$, $\phi_{p'}$ coincides with $\phi$.

Since $x$ cannot be experimentally accessed, GPDs cannot be directly measured. 
Some help comes from the fact that
the 
leptoproduction of a real photon always occurs through two different mechanisms leading to the same final state $(e' \gamma p')$: the DVCS process, discussed above and related to the parton content of the target, and the electromagnetic Bethe-Heitler (BH) process, shown in Fig. 2. In facts, the complete squared amplitude for the leptoproduction process has to be read as 
\begin{equation}\label{amp}
    \mathcal{A}^2 = T_{DVCS}^2+T_{BH}^2 + \mathcal{I}\, .
\end{equation}
In particular, in the kinematical region tested at JLab and 
of interest here,
the BH mechanism is dominating the DVCS one. For this reason, a key handle to access the GPDs is the interference between these two competing processes, i.e. $ \mathcal{I} = 2 \Re e(T_{DVCS}T_{BH}^*)$. This term, containing $T_{DVCS}$ is sensitive to the parton content of the target through the GPDs. Such information is encapsulated in the Compton Form Factors (CFFs) $\mathcal{F}$ related to the generic GPDs $F$ by:
\begin{equation}\label{ccfegpd}
\mathcal{F}(\xi,\Delta^2) = \int dx \frac{F(x,\xi,\Delta^2)}{x\pm \xi + i \epsilon}\,.
\end{equation}
Since in the CFFs the dependence on $x$ is integrated out, they can be measured.
Also for the CFFs the obvious $Q^2$ dependence is omitted here and in the following.
We note in passing that the possibility that the final photon is emitted by the initial nucleus, or by the final nuclear system X, has 
been neglected, being the BH cross section approximately 
proportional to the inverse squared mass of the emitter. Therefore, with respect to the emission from the electrons, this contribution is negligibly small. In facts, the experimental collaboration EG6 has not considered this occurrence in its analysis. From a theoretical point of view, if these contributions are neglected, gauge invariance is not
respected. Nonetheless, we have to point out that in the 
present IA analysis gauge invariance is 
in any case not fulfilled and it could be restored only implementing many-body currents at the nuclear level. These corrections have not been included in the calculation yet and they could be more relevant than photon emission from 
nuclear systems in the initial and final state.

The clearest way to experimentally access the relevant interference term is the measurement of the beam-spin asymmetry (BSA) for the process where the unpolarized target (U), $^4$He in this case, is hit by  a longitudinally polarized (L) electron beam with different helicities ($\lambda = \pm$). So, the observable under scrutiny reads
\begin{equation}\label{ALU}
    A_{LU}= \frac{d\sigma^+ - d \sigma^-}{d \sigma^+ + d \sigma^-}.
\end{equation}

Since the interference term is directly proportional to the helicity of the beam, the difference of cross sections for different beam helicities in the numerator of Eq.\eqref{ALU},
up to a phase space factor, gives a direct access to such term.
We will show in the following that the quantities
$d \sigma^\pm$ in Eq. \eqref{ALU}
are actually 4-times differential cross sections.


{Our aim is thus the evaluation of the complete expression for the leptoproduction cross section at LO in IA in order to study the theoretical behaviour of the BSA
and compare it with the data. The details of the calculation are described in the following.}

\begin{figure}[hpt]
\hspace{-.5cm}
\includegraphics[scale=0.3,angle=0]{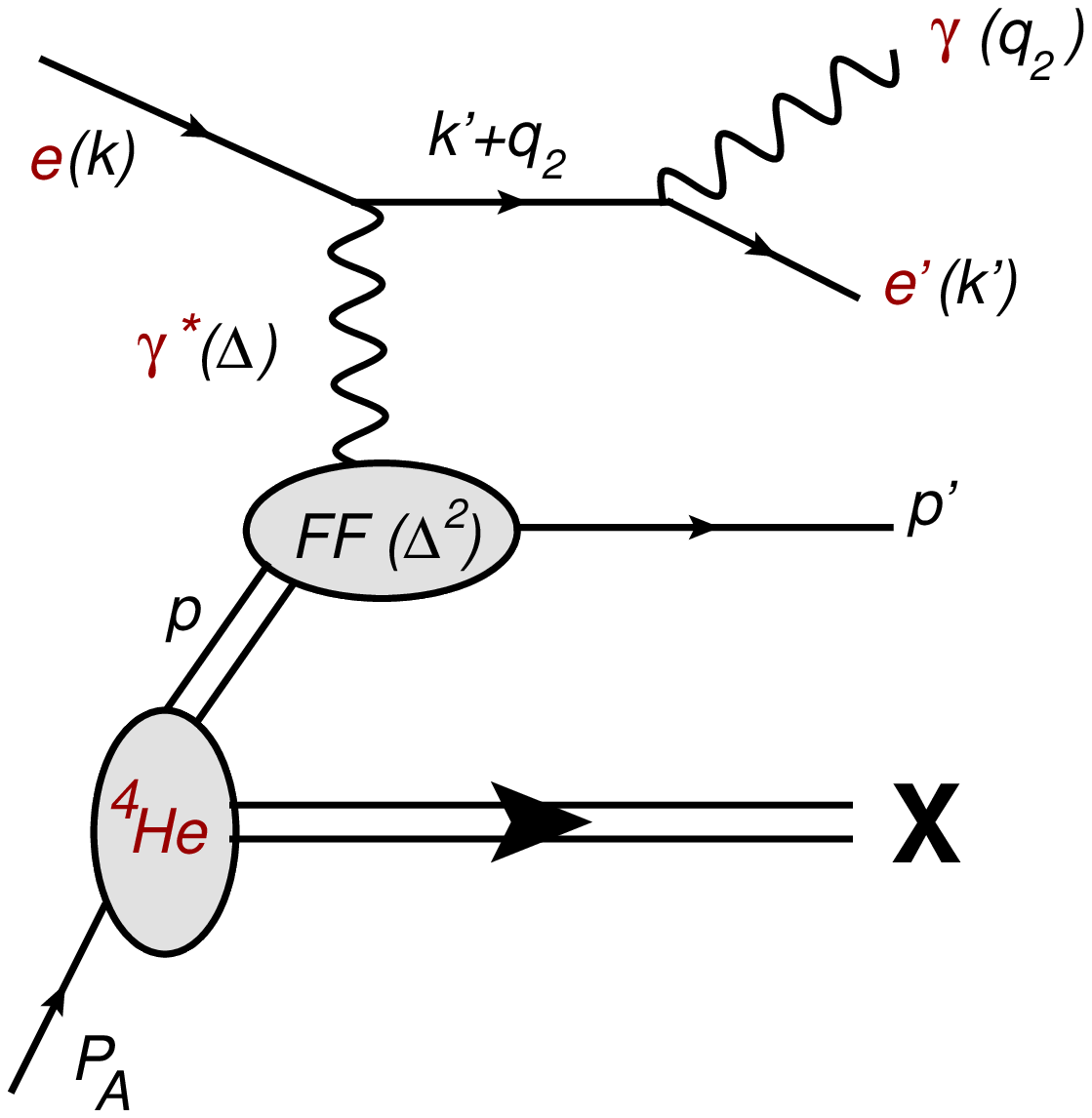}
\hspace{0.6cm}
\includegraphics[scale=0.3,angle=0]{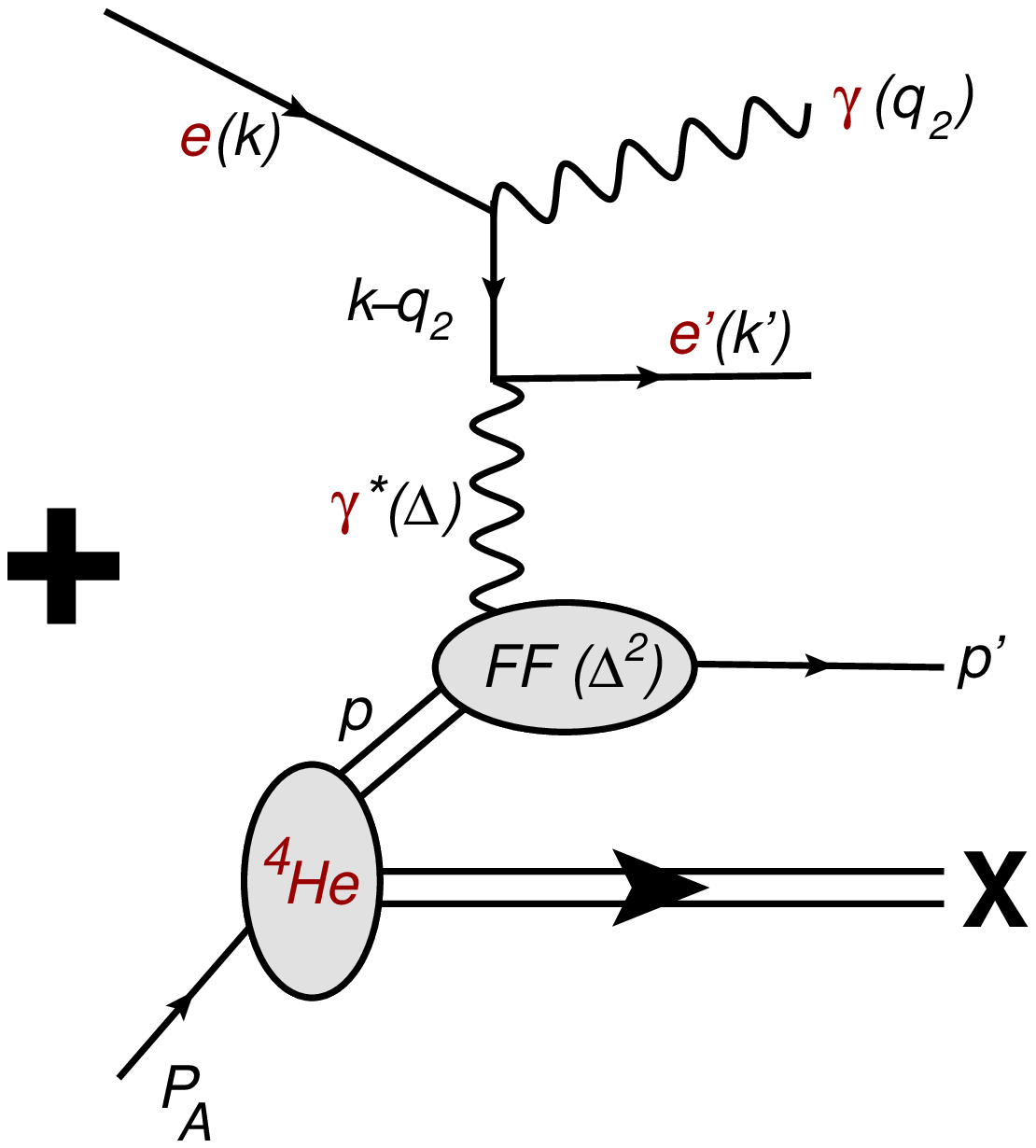}
\caption{(color online) The Bethe Heitler process in IA.}
\label{dvcshb}
\end{figure}
In our IA approach, {we account only for the kinematical off-shellness of the initial bound proton so that}
the energy of the struck proton
is obtained from energy conservation and reads
\begin{equation}
    p_0 = M_A  - \sqrt{M_{A-1}^{*2} + \vec{p}^2}\simeq M - E - T_{rec}\,, 
\label{off}
\end{equation}
where we define the removal energy $E = M^*_{A-1} + M - M_A = \epsilon^*_{A-1}+ |E_A| - |E_{A-1}|$  in terms of the binding energy (mass) of $^4$He and of the 3-body system, $E_A$ ($M_A$) and $E_{A-1}$
($M_{A-1}^*$), respectively, and of the excitation energy
of the recoiling system, $\epsilon^*_{A-1}$. Finally,  $T_{rec}$ is the kinetic energy of the recoiling  $3-$body system and $M$ is the proton mass.
A straightforward but lengthy analysis,
detailed in appendix \ref{convfor}, leads to a
complicated convolution formula for the cross section, which can be cast in the following form
\begin{equation}\label{cross}
    d\sigma_{Inc}^\pm= 
    \int_{exp} dE \, d{\vec p} \,
    \frac{p\cdot k}{p_0 \, |\vec k|}
 P^{^4He}(\vec{p},E)\, d\sigma^{\pm}_b(\vec p, E,K)\,,
\end{equation}
where the main ingredients are the nuclear spectral function $ P^{^4He}(\vec{p},E) $ and the cross section for a DVCS process off a bound proton,
$d\sigma^{\pm}_b$. 
As thoroughly described in Appendix A,
the integral on the removal energy refers to 
the full spectrum of $^4$He,
both discrete and continuous.
In Eq. \eqref{cross}, $K$ is the set of kinematical variables $\{x_B=Q^2/(2 M \nu),Q^2,t,\phi\}$.
The range of $K$ accessed
in the experiment fixes the proper
energy and momentum
integration space, denoted
as $exp$ and described in
appendix \ref{convfor}.
From Eq. \eqref{cross} we get
the measured differential cross sections, 
appearing in Eq. \eqref{ALU},
\begin{eqnarray}
\label{crossl}
d \sigma^\pm \equiv  \frac{d\sigma^\pm_{Inc} }{d{x}_BdQ^2 d{\Delta}^2 d\phi}&= \int_{exp} dE \, d{\vec p} \,P^{^4He}(\vec {p},E)
\\
  & \times |\mathcal{A}^{\pm}({\vec p}, E ,K)|^2
g(\vec{p},E,K) \nonumber \,,
\end{eqnarray}
where  $g(\vec p,E,K)$ is a complicated function
which arises, as explicitely detailed in Appendix A, from the integration over the phase space and includes also the flux factor ${p\cdot k}/({p_0 \,|\vec  k |})$ of Eq. \eqref{cross}. 
This latter term comes from the fact that one has at disposal only  non-relativistic nuclear wave functions to evaluate the spectral function. In the present approach
this implies that the number
of particle sum rule is respected, but the momentum sum rule is slightly
violated. Such a  problem could be solved ultimately within a Light Front approach, along the lines proposed in Ref. \cite{DelDotto:2016vkh} for a $3-$body system.
 
The BSA \eqref{ALU}, written in terms
of the above cross sections, yields the schematic form
\begin{equation}\label{alu_ratio}
A_{LU}^{Incoh} (K) = 
 \frac
 { \mathcal{I}^{^4He} (K) } 
 { T_{BH}^{2\, \, ^4He} (K) 
 }\,,
\end{equation}
where
\begin{equation}\label{nume}
\mathcal{I}^{^4He} (K) =
\int_{exp} dE \, d \vec p \, 
{ {P^{^4He}(\vec p, E )}} 
\, g(\vec p,E,K)\, 
{{\mathcal{I}(\vec p, E, K)}} \,,
\nonumber
\end{equation}
\begin{eqnarray}\label{deno}
T_{BH}^{2\,\,^4He} (K) & = &
\int_{exp} dE \, d \vec p \, {{P^{^4He} (\vec p, E )}} 
\, g(\vec p, E,K)\,\nonumber \\ 
& \times &
{{T_{BH}^2 (\vec p, E, K)}}\,,
\end{eqnarray}
refer to a moving bound nucleon and
generalize the Fourier decomposition of the DVCS cross section off a proton at rest, at leading twist, derived in
Ref. \cite{Belitsky:2001ns}.
Without going into technical details, that are presented in  appendix \ref{scattamp}, we summarize 
the structure of the different contributions.\\
For the BH part, we considered the full sum of azimuthal harmonics, i.e
\begin{equation}
   T _{BH}^2 = c_0^{bound} + c_1^{bound}\cos \phi + c_2^{bound} \cos(2 \phi )\, ,
\end{equation}
where the coefficients $c_i^{bound}$ contain the Dirac and Pauli form factors (FFs).
{The azimuthal dependence of the amplitudes is due to the expression of the BH propagator as reported in Appendix \ref{scattamp}. }
We stress that in the present IA approach no nuclear modifications occur for the FFs of the bound proton.
Concerning the interference part {in the numerator of Eq. \eqref{ALU}}, 
terms proportional to $\Delta^2/Q^2$ 
have been considered as well as corrections proportional to  $\epsilon^2 =4 M^2 x_B^2/{Q^2}$, accounting for target mass corrections. The latter terms are fundamental in order to obtain a fully consistent comparison with the BSA for a proton at rest, which will be shown in the next section. The main reason is that in the amplitudes for a bound proton it is not always possible to isolate such terms, since the obtained expressions are function of the 4-momentum of the bound, off-shell proton.
In our approach
the parton content of the bound proton 
plays a role only 
in the imaginary part of the CFF $\mathcal{H}$. 
In the kinematics of interest { and in the present model}, this quantity can be 
expressed in terms of only one GPD 
of the bound proton, $H(x,\xi,\Delta^2)$,  
selected in the slice $ x = \pm \xi$ , i.e.
\begin{equation}
\Im m \mathcal{H}(\xi',t) = H(\xi',\xi',t)- H(-\xi', \xi', t)\, ,
\end{equation}
{where $H(\xi',\xi',t)$ is summed over the $u, d, s$ flavours of the quarks.}
We notice that the off-shellness of the bound nucleon
enters the proton parton structure through the dependence of the GPDs on $\xi' = {-q^2}/({P \cdot q})$. In this way, 
the modification at partonic level is due to this rescaling of the skewness that, for a proton at rest, becomes $ \xi = x_B(1+\Delta^2/2 Q^2)/(2 - x_B +x_B\Delta^2/Q^2) $, keeping terms proportional to $\Delta^2/Q^2$.

\begin{figure}
    \centering
\includegraphics[scale=0.6]{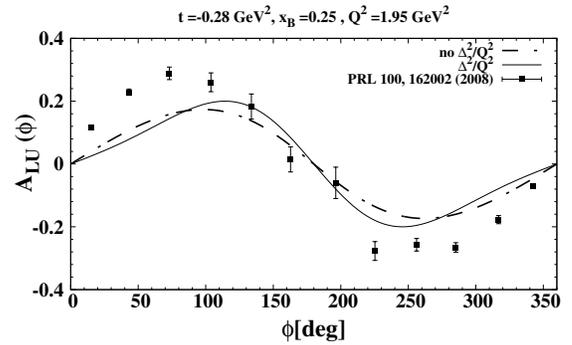}
    \caption{Beam spin asymmetry for a proton at rest considering (full curve) and ignoring (dot-dashed curve) term of order $\Delta^2/Q^2$ in the interference part. In this kinematics,  $\Delta^2/Q^2 \simeq 0.144$.
    Data from Ref. \cite{Girod:2007aa}.}
    \label{girod}
\end{figure}

\begin{figure}[htbp]
\hspace{-.5cm}
\includegraphics[scale=0.6,angle=0]
{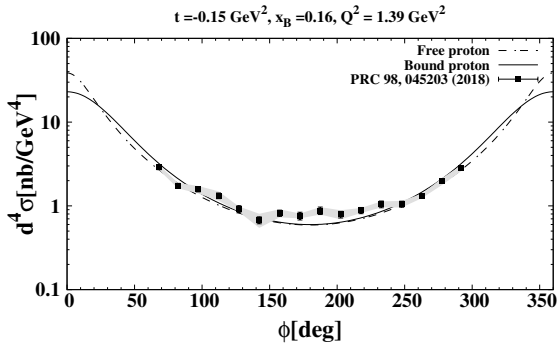} 
\caption{The cross section for the BH process
on the free proton (dashed line) and on a proton bound
in $^4$He (full line), according to the present treatment, in the kinematics reported on the top
of the frame, corresponding to data presented in
Ref. \cite{HirlingerSaylor:2018bnu}, as a function of the azimuthal angle
$\boldsymbol{\phi}$. The precise position of the data and their errors are taken from \cite{CLAS}.}
\label{motfer}
\end{figure}

\begin{figure}[htbp]
\hspace{-.5cm}
\includegraphics[scale=0.6,angle=0]{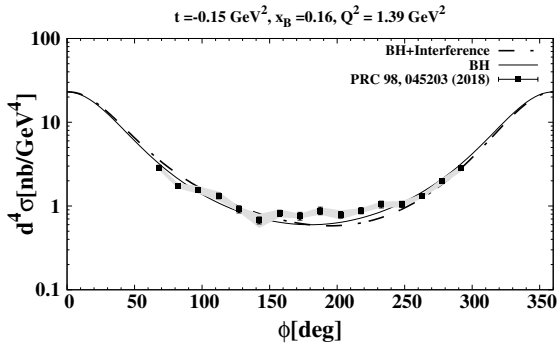} 
\caption{
The cross section for the BH process (full line)
and the one obtained including the interference between the BH and DVCS processes (dot-dashed line),
for a proton bound
in $^4$He, according to the present treatment, in the kinematics reported on the top
of the frame, corresponding to data presented in
Ref. \cite{HirlingerSaylor:2018bnu}, as a function of the azimuthal angle
$\boldsymbol{\phi}$. The precise position of the experimental data and their errors are taken from \cite{CLAS}.
}
\label{cross?}
\end{figure}

\begin{figure}[htbp]
\hspace{-.5cm}
\includegraphics[scale=0.6,angle=0]{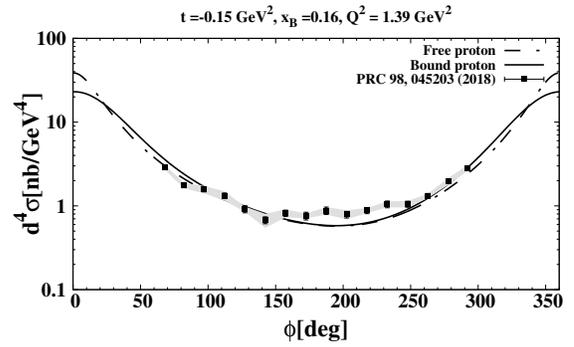}
\caption{
The cross section for the bound proton (full line) and for the free proton (dot-dashed line) in the kinematics reported on the top
of the frame, corresponding to data presented
in Ref. \cite{HirlingerSaylor:2018bnu}, as a function of the azimuthal angle
$\boldsymbol{\phi}$. 
The precise position of the data and their errors are taken from \cite{CLAS}.}
\label{cross_bh}
\end{figure}

\begin{figure}[htbp]
\hspace{-.5cm}
\includegraphics[scale=0.6,angle=0]{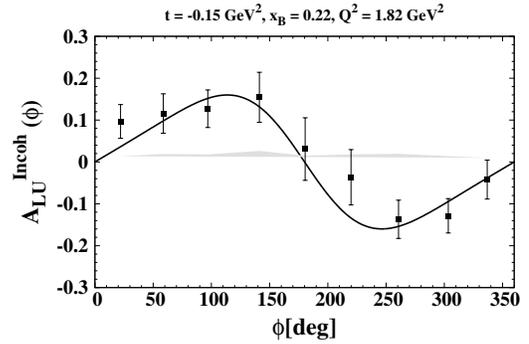} 
\caption{The BSA 
$A_{LU}^{Incoh}$,
Eq. \eqref{alu_ratio},
as a function of the azimuthal ange $\boldsymbol{\phi}$,
compared to data corresponding to the analysis
leading to Ref \cite{Hattawy:2018liu}
}
\label{aluphi}
\end{figure}

\begin{figure*}[htbp]
    \centering
    \includegraphics[scale=0.85]{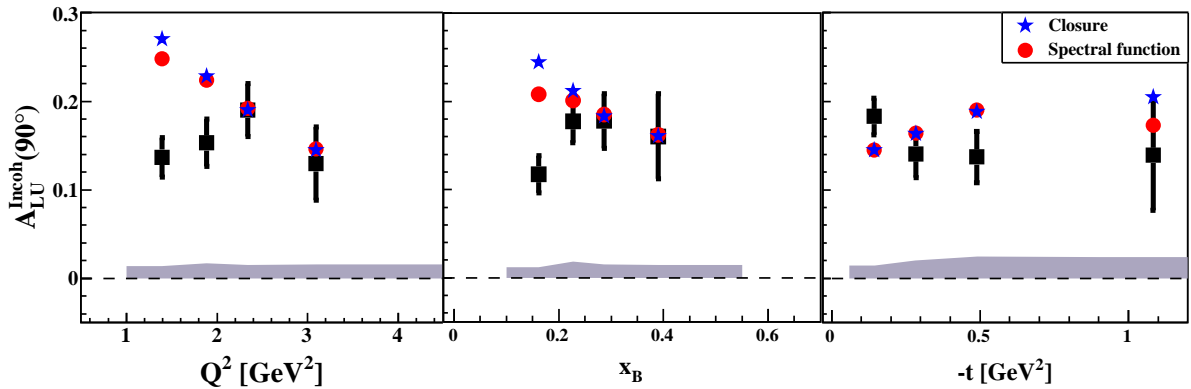}
5   \caption{(Color online)
Azimuthal beam-spin asymmetry for the proton in
$^4$He, $A_{LU}^{Incoh}(K)$,
for $\phi = 90^o$: results of this approach (red dots are obtained 
using the  diagonal spectral function as described along the text, 
blue stars using the momentum distribution in the so called  ''closure approximation"
) compared with data
(black squares)
\cite{Hattawy:2017woc}.
From left to right, the quantity is shown in the experimental
$Q^2$, $x_B$ and $t$ bins, respectively. 
Shaded areas represent systematic errors.}
   \label{closure}
\end{figure*}

\begin{figure*}[htbp]
\hspace{-.5cm}
\includegraphics[scale=0.85,angle=0]{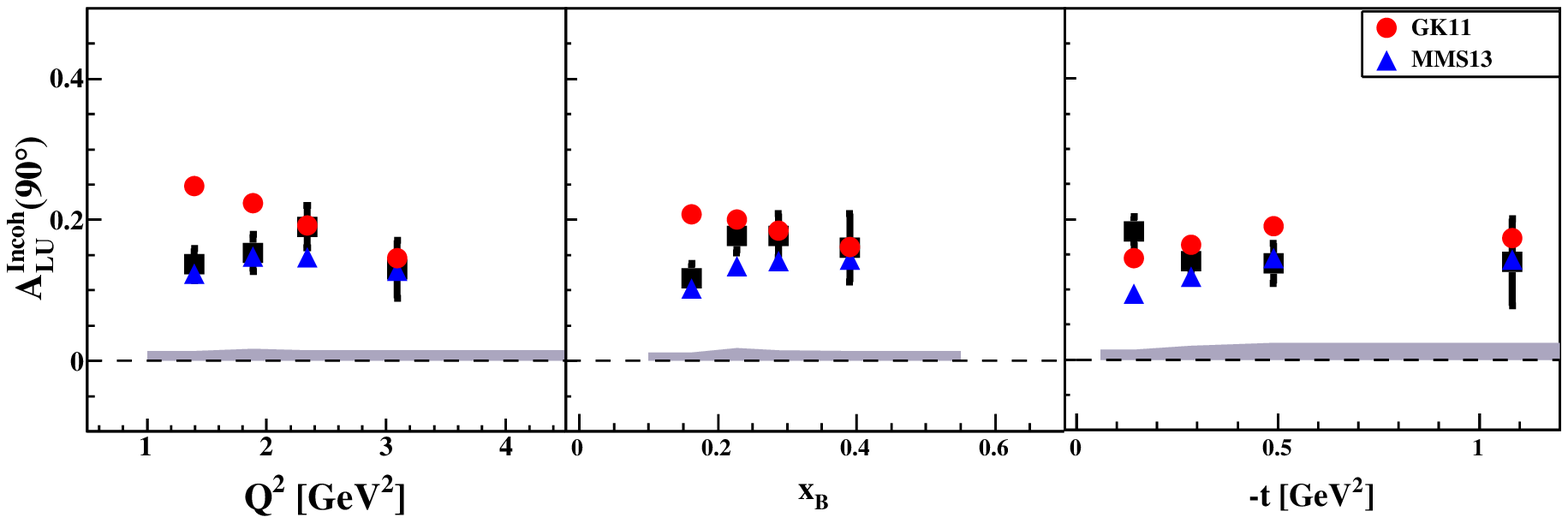}
\caption{
(Color online)
Azimuthal beam-spin asymmetry for the proton in
$^4$He, $A_{LU}^{Incoh}(K)$,
for $\phi = 90^o$: results of this approach (red dots are obtained 
using the GK GPD model
\cite{Goloskokov:2011rd}, blue triangles using the MMS
model
\cite{Mezrag:2013mya}
) compared with data
(black squares)
\cite{Hattawy:2017woc}.
From left to right, the quantity is shown in the experimental
$Q^2$, $x_B$ and $t$ bins, respectively. 
Shaded areas represent systematic errors.
}
\label{aluexp}
\end{figure*}

\section{Ingredients of the calculation}

In order to actually
evaluate Eq. \eqref{alu_ratio},
we need an input
for the proton GPD and for the proton spectral function
in $^4$He. 
Concerning the nuclear part,
only old attempts exist of obtaining a complete spectral function of $^4$He
\cite{Morita:1991ka,trento}.
The unpolarized spectral function, whose emergence in this process is thoroughly
described in appendix A, can be cast in the form
\begin{eqnarray}
P^{^4He}(\vec p, E)
& = & \sum_{f_{A-1}}
\langle {^4He}|f_{A-1};
N \,\vec p \,
\rangle \langle
f_{A-1};
N \,\vec p \, | 
{^4He} \rangle
\nonumber
\\
& \times & 
\delta(E-E_{min}-\epsilon_{A-1}^*) \, .
\label{overl}
\end{eqnarray}
It is therefore clear that
its realistic evaluation would require
the knowledge, at the same time, of exact solutions
of the Schr\"odinger equation with realistic nucleon-nucleon potentials and three-body forces
for the $^4$He nucleus and for the three-body recoiling system $f_{A-1}$.
This system can be either in its ground state, when
$E=E_{min}=|E_{^4He}| - |E_{^3H}|$,
or unbound with an excitation
energy $\epsilon^*_{A-1}$.
The description of this latter part represents a challenging
few-body problem, whose solution
is presently unknown. 
A full realistic calculation
of the $^4$He spectral function is planned
and has started but,
in this work, for
$P^{^4He}(\vec p, E)$
use is made of the model 
presented in Ref. \cite{Viviani:2001wu,Rinat:2004ia}.
In that approach,
when the recoiling system is in its ground state
and $E=E_{min}$,
an exact description is used
in terms of
variational wave functions
for the 4-body 
\cite{PisaWF} 
and 3-body 
\cite{PisaWF3} 
systems,
obtained through the hyperspherical 
harmonics method \cite{hh},
within the Av18 NN interaction \cite{Wiringa:1994wb},
including UIX three-body forces \cite{Pudliner:1995wk}.
The cumbersome part of the spectral function, with the recoiling system excited,
is  
based on the Av18+UIX interaction, proposed
in Ref. \cite{Viviani:2001wu,Rinat:2004ia},
an
update of the two-nucleon correlation model of Ref. \cite{CiofidegliAtti:1995qe}.
We note that realistic calculations of GPDs for $^3$He,
for which an exact spectral function is available, have
established the importance of properly
considering the $E$-dependence
of the spectral function \cite{Scopetta:2009sn}.
To have an idea of the importance of a proper treatment of the $E$-dependence in this process, and, in general, of the drawback of the use of a less refined nuclear description, in the next section we will show also results obtained using the so called "closure" approximation.
It consists in 
evaluating the spectral function considering,
in the argument of the delta function in Eq. \eqref{overl},
an average value of the removal energy, so that
the closure of the $f_{A-1}$ states can be used, yielding
\begin{eqnarray}
P^{^4He}_{closure}(p, E) & = &
n_{gr}(p) \delta(E-E_{min})
\nonumber
\\
& + &
n_{ex}(p) \delta( E - \bar E) \, ,
\label{clos}
\end{eqnarray}
where the momentum distribution for the proton
with the recoiling system in its ground or excited state,
$n_{gr}(k)$ and $n_{ex}(k)$, respectively, have been
introduced, with $\bar E$ the average excitation energy
of the recoiling system.
A similar approach has been used to model the non-diagonal
$^4$He spectral function in the description of coherent DVCS
off $^4$He, in Ref. \cite{Fucini:2018gso}.
We note that, when this approximation is used, also
the off-shellness of the struck proton, governed
by Eq. 
\eqref{clos}, 
has to be changed accordingly, i.e.
\begin{equation}
    p_0 = M_A  - \sqrt{M_{A-1}^{*2} + \vec{p}^2} \longrightarrow M - \bar E - T_{rec}\,.
\label{offnew}
\end{equation}

As we will see in the following,
this produces important effects in the cross section, 
due to the fact that the components of the four momentum
of the proton enter scalar products present
in the relevant scattering amplitudes.

For the nucleonic GPD, two models have been used.
One is the model of
Goloskokov and Kroll (GK) 
\cite{Goloskokov:2011rd}, already
successfully exploited in the coherent case
\cite{Fucini:2018gso}. It is worth
to remind that
the model is valid in principle at $Q^2$ values larger than those of interest here, in particular at $Q^2 \ge 4$ GeV$^2$.
Nonetheless we have 
checked that the GK model can reasonably describe free
proton data collected in similar kinematical ranges, for example the ones in Ref. \cite{Girod:2007aa},
as it is discussed in the next section
(see Fig. \ref{girod}).

The other model is taken from Ref. \cite{Mezrag:2013mya}. It is
based on an original compact version of the double distribution 
prescription.
It is developed at leading twist
and at leading order in $\alpha_s$ 
(of course NLO corrections may be sizable also in the valence region, at moderate energy, see, e.g., the discussion
in Ref. \cite{Moutarde:2013qs}).
With respect to the GK model, 
only the valence region is modified 
and the momentum scale evolution is the same.
The model is expected to work
in the region $-t/Q^2\leq0.1$, where  factorization is supposed to work.
To obtain the relevant numbers for that model, use has been made
of the virtual access infrastructure "3DPARTONS"
\cite{Berthou:2015oaw}.


\section{Numerical results}

We can now evaluate 
the beam spin symmetry (BSA),
Eq. \eqref{alu_ratio} ,
and compare it with the recently published data \cite{Hattawy:2018liu}.

First of all, let us check if the GK
model we used, for values of
$Q^2$ smaller than those for which
it is supposed to work, $Q^2 \ge 4$ GeV$^2$,
is still describing the available data
reasonably well.
To this aim, we show in Fig. \ref{girod} that,
in one of the kinematics presented in Ref.
\cite{Girod:2007aa} for DVCS off the free proton,
not far from the ones of interest here,
a reasonable description of the BSA data, is obtained
calculating this quantity for the free proton with the GK model.
{We notice that the azimuthal angle $\boldsymbol{\phi}$, used by the experimental collaboration and
exploited here, is related to the one previously defined and used
in this paper by the relation $\boldsymbol{\phi}= \pi - \phi$.
}
The relevance of terms of order $t/Q^2$, discussed in the previous section,
is also shown. In general, the BSA is rather sensitive
to changes of the kinematics, to $t$ especially.
Data for the free proton are not available for the
kinematics of the experiment under scrutiny
so that we have to compare with results of other experiments.

Then, let us show the results of our model for the differential cross sections \eqref{crossl} which are used later to calculate the BSA.
All the cross sections shown here below are obtained
considering a positive electron helicity, as an example.

To have a first glimpse at the nuclear effects on the relevant processes,
the cross section for the BH process
on the free proton (dashed) and on a proton bound
in $^4$He (full), according to the present treatment, is shown in Fig. \ref{motfer},
as a function of the azimuthal angle $\boldsymbol{\phi}$,
in one of the kinematical ranges 
of the data presented in
Ref. \cite{HirlingerSaylor:2018bnu}. 
The data, corresponding
to the full DVCS process off the free proton, are presented here for illustration only.
Relevant nuclear effects are clearly seen.
To our knowledge, this figure and the next two are the
first ones in the literature where the
comparison of cross sections for free and bound
nucleons, with a difference arising from
a microscopic calculation, is presented.

In Fig. \ref{cross?},
the cross section for the BH process
is compared with that obtained including also the only relevant term, as discussed in Appendix B, of the
the interference between the BH and DVCS processes,
for a proton bound
in $^4$He according to the present treatment, again
in the kinematics of
Ref. \cite{HirlingerSaylor:2018bnu}, as a function of the azimuthal angle
$\boldsymbol{\phi}$ (see Appendix B for the discussion of the relevant term included).
It is clearly seen as a relevant $\phi$ asymmetry is generated
including the DVCS mechanism. 
The data for the free proton are again reported for illustration.
It is seen that a reasonable description is obtained.

In Fig. \ref{cross_bh}, in the same kinematics 
of the previous two, the full cross section
is shown, for a bound and for a free proton, 
to expose the role
of the nuclear effects on the proton DVCS cross-section,
found to be overall sizable.


Let us now present results for
the BSA 
$A_{LU}^{Incoh}$,
Eq. \eqref{alu_ratio}.
This quantity, evaluated
using the GK model for the GPD entering
the DVCS part,
is shown in Fig. \ref{aluphi},
as a function of the azimuthal ange $\boldsymbol{\phi}$,
compared to data corresponding to the analysis
leading to Ref \cite{Hattawy:2018liu}.
A convincing agreement is found, in particular
at $\boldsymbol{\phi}=\phi=90^o$, the fixed value at which
the BSA has been extracted and
at which it will be shown in the following.

The BSA is a function of the azimuthal angle $\phi$
and of the kinematical variables $Q^2$, $x_B$ and $t$. Due to limited 
statistics, in the experimental analysis
these latter variables have been studied separately 
with a two-dimensional 
data binning. The same procedure has been used in our calculation.
For example, each point at a given $x_B$ has been obtained
using for $t$ and $Q^2$ the corresponding average experimental values, which are reported for definiteness in Tables I-III,
together with the numerical values of the calculated theoretical
asymmetries discussed in the following.

\begin{table}[ht]
    \centering

\begin{tabular}{|c|c|c|c|c|}
\hline
$x_B$ & $ <Q^2>$ [GeV$^2$]  &  $ <t>$ [GeV$^2$]   &  $A_{LU}^{GK}$ & $A_{LU}^{MMS}$ \\
     \hline
   0.162 & 1.43  &  -0.397 &  0.208 & 0.102\\
      \hline
  0.227& 1.92& -0.418  &  0.204 & 0.134\\
      \hline
  0.287 & 2.35 & -0.492   & 0.185 & 0.141\\
      \hline
   0.390 & 2.98 & -0.714 & 0.163 & 0.143\\
     \hline

\end{tabular}

\caption{The BSA, obtained using the GK 
\cite{Goloskokov:2011rd}
or MMS \cite{Mezrag:2013mya}
models, using the nuclear spectral function, for the average values of $Q^2$ and $t$ in $x_B$ bins.}

\end{table}

\begin{table}[hbpt]
    \centering

\begin{tabular}{|c|c|c|c|c|}
\hline
  $Q^2$ [GeV$^2$] & $ <x_B>$ &  $ <t> $ [GeV$^2$]  & $A_{LU}^{GK}$ &  $A_{LU}^{MMS}$ \\ \hline
   1.40 & 0.166  &   -0.407 & 0.248 & 0.124\\
      \hline
   1.89 & 0.233  &  -0.499  & 0.224 & 0.148\\
      \hline
   2.34 & 0.290  & -0.521  & 0.192 & 0.147\\
      \hline
   3.10 &  0.379 & -0.650 & 0.146 & 0.128\\
     \hline
     
\end{tabular}
\caption{The BSA, obtained using the GK 
\cite{Goloskokov:2011rd}
or MMS \cite{Mezrag:2013mya}
models, using the nuclear spectral function,
for the average values of $x_B$ and $t$ in $Q^2$ bins.}
\end{table}

\begin{table}[hbpt]
    \centering

\begin{tabular}{|c|c|c|c|c|c|c|}
\hline
     $t$ [GeV$^2$]& $ <x_B>$ &  $ <Q^2>$ [GeV$^2$] & $A_{LU}^{GK}$ &  $A_{LU}^{MMS}$ \\
   \hline 
   -0.145 &  0.213 & 1.82  & 0.145 & 0.094\\
      \hline
   -0.282 & 0.255 &2.13   &  0.164 & 0.118 \\
      \hline
   -0.490 & 0.284  &  2.31 & 0.190 & 0.144\\
      \hline
   -1.11 & 0.308 &  2.41 & 0.173 & 0.140\\
     \hline

\end{tabular}

\caption{The BSA, obtained using the GK 
\cite{Goloskokov:2011rd}
or MMS \cite{Mezrag:2013mya}
models, using the nuclear spectral function, 
for the average values of $x_B$ and $Q^2$ in $t$ bins.}
\end{table}


In Fig.~\ref{closure} it is seen that,
overall, the calculation reproduces the data
rather well in all of these bins.
For this observable, in most of the cases
the present accuracy of the data does not allow
to distinguish between the full calculation and
that performed using the closure approximation,
Eq. \eqref{clos}. In any case, whenever the disagreement with the data
is sizable, the proper treatment of the excitation energy within the spectral function helps in describing the data.
Besides, we note that the agreement is not satisfactory only when the GK model is used
in the region
of low $Q^2$.
Indeed, this is evident only in the experimental
points corresponding to the lowest values
of $Q^2$, $x_B$ and $t$.
One should notice that the average value
of $Q^2$ grows with increasing $x_B$
and $t$ (cf. tables I-III), so that a not satisfactory description
at low $Q^2$ affects also the first $x_B$
and $t$ bins. Actually, the GK model is designed to describe
the available data for $Q^2 \le 4$ GeV$^2$, e.g at
values higher than the typical ones accessed by the CLAS collaboration in the experiment under scrutiny.
The problems found using the GK parametrization
are therefore somehow expected.
We have therefore repeated the calculation using
as a nucleonic partonic input the model MMS,
introduced in Ref. \cite{Mezrag:2013mya},
briefly described in the previous section.
The comparison of
the two results is presented in Fig \ref{aluexp},
where it is seen that the data favor the MMS model 
with respect to the GK one.
The success of the MMS model, with parameters chosen precisely to be realistic in the $Q^2$ range typical at JLab, is remarkable and points to a solid predictivity of the IA, emphasizing, at the same time, the dependence of the results on the choice of the nucleonic model. 
In any case, the residual disagreement, or the
problems found using the GK model, could be also due to some final state interaction (FSI) effects that in the present IA are not considered. For this reason, a careful analysis of the interplay between
the $t$ and $Q^2$ dependence of the data
is required to establish whether FSI play a relevant role. The present accuracy of the data
does not allow such an analysis, but the data expected
from the planned future measurements certainly will.
In the light of this discussion, we can conclude that
a careful use of basic conventional 
ingredients is able to reproduce the available
data.
In order to better understand our results, addressing nuclear modifications of the parton strucure,
possibly related therefore to the EMC effect, as an illustration we perform a specific analysis, detailed
in what follows.

Let us define, in each experimental bin, specific ratios
to expose the nature of nuclear effects, namely,
the ratio between the BH-DVCS interference cross section
for the proton bound in $^4$He and the free one at rest,
$R_{\mathcal{I}}(K)$, the corresponding quantity
for the pure $BH$ process,   $R_{{BH}}(K)$,
and the ratio of the two, $R_{ALU}(k)$, providing the ratio 
of the bound proton to the free proton
$BSA$
in our calculation scheme.
These quantities read, respectively
\begin{eqnarray}
    R_{\mathcal{I}}(K) & = &\frac{1}{\mathcal{N}}
    \frac{ {I}^{^4He} (K)}{{I}^{p} (K)} \, ,
    \label{r1}
    \\
  R_{BH} (K) &= &\frac{1}{\mathcal{N}}   
    \frac{T_{BH}^{2\,\,^4He} (K) }{T_{BH}^{2\,p} (K) } \, ,
\label{r2}
    \\
    R_{ALU} (K) & = & \frac{R_\mathcal{I}(K)} { R_{BH} (K) } 
    = \frac{ A_{LU}^{Incoh} (K) } { A_{LU}^p (K) }
    \, .
\label{r3}
\end{eqnarray} 
\begin{figure*}[htbp]
\hspace{-.5cm}
\includegraphics[scale=0.85,angle=0]{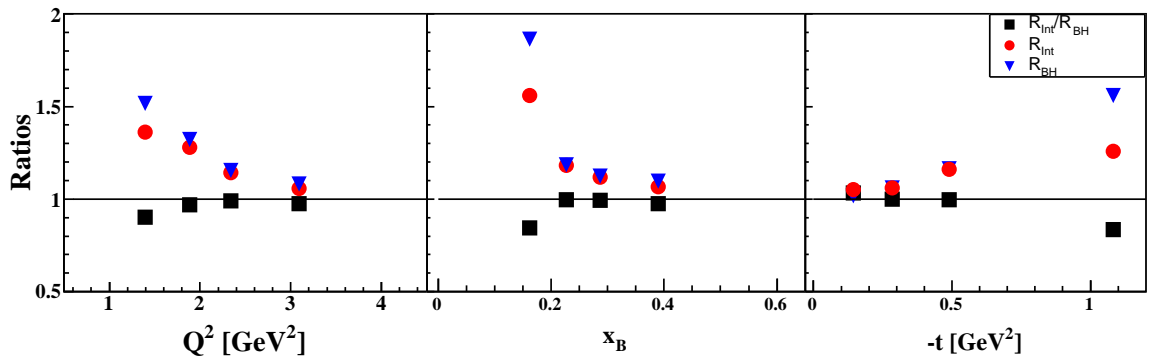}
\caption{
(Color online)
The ratios \eqref{r1} (red dots), \eqref{r2} (blue triangles), \eqref{r3} (black squares), at 
$\phi = 90^o$ and using the GK model for the nucleon GPD. 
From left to right, the quantity is shown in the experimental
$Q^2$, $x_B$ and $t$ bins, respectively.
}
\label{ratiosep}
\end{figure*}
In the equations above the factor
${\cal N}=\int_{exp} dE \, d\vec{p} P^{^4 He} ( \vec{p}, E)$
accounts for the 
fact that only a part of the
spectral function is selected in a given
experimental bin.
The meaning of the integration space $exp$ is clarified in appendix $A$.
The ratios \eqref{r1}-\eqref{r3} at 
$\phi = 90^o$, using the GK model for the nucleon GPD,
are shown in Fig. \ref{ratiosep}.
It is clearly seen that the nuclear effects obtained 
within the present IA scheme in the
ratios \eqref{r1} and \eqref{r2} are rather sizable,
while the effects are dramatically reduced in the "super-ratio"
\eqref{r3}. This fact points to relevant conventional
nuclear effects in the pure BH and pure DVCS processes, which are anyhow
of a similar origin, so that they cancel out to a large extent
in the ratio.

\begin{figure*}[htbp]
\hspace{-.5cm}
\includegraphics[scale=0.85,angle=0]{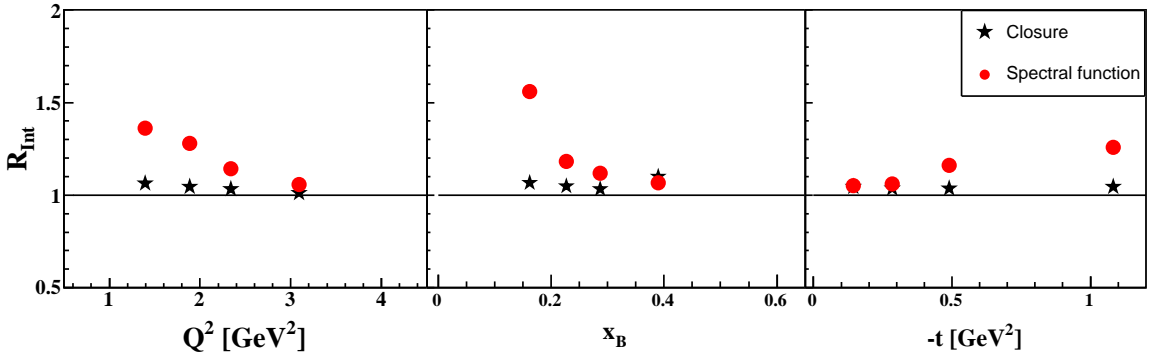}
\caption{
(Color online)
The ratio \eqref{r1} (blue triangles), obtained using either the spectral function (red dots) or the closure approximation (black stars), at 
$\phi = 90^o$ and using the GK model for the nucleon GPD. 
From left to right, the quantity is shown in the experimental
$Q^2$, $x_B$ and $t$ bins, respectively.
}
\label{intclo}
\end{figure*}

\begin{figure*}[htbp]
\hspace{-.5cm}
\includegraphics[scale=0.85,angle=0]{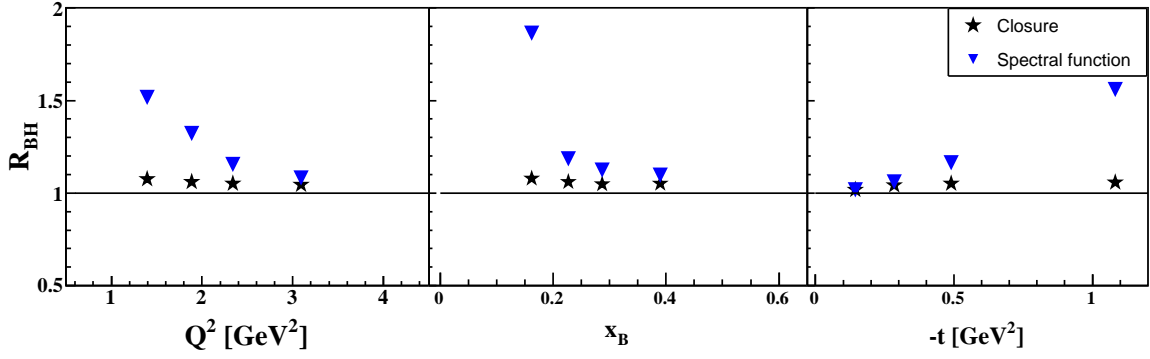}
\caption{
(Color online)
The ratio \eqref{r2}, evaluated using the spectral function (blue triangles) and the closure approximation (black stars) for
$\phi = 90^o$ and using the GK model
for the proton GPD.
From left to right, the quantity is shown in the experimental
$Q^2$, $x_B$ and $t$ bins, respectively.
}
\label{bhclo}
\end{figure*}

Something similar happens when the closure approximation
is applied to estimate the nuclear effects.
In Figs. \ref{intclo} and \ref{bhclo} it is seen
that, in some cases, the difference between
the results of the full calculation, performed considering the distribution
of the removal energy within the spectral function,
and of the one obtained with the closure approximation,
is rather sizable in the ratio \eqref{r1} and \eqref{r3}.
In Fig. \ref{ratioexp} is seen instead that
the effect is dramatically reduced in the ratio
of these two quantities, the super-ratio
\eqref{r3}, showing that
the effects in the numerator and in the denominator basically compensate each other. 

\begin{figure*}[htbp]
\hspace{-.5cm}
\includegraphics[scale=0.85,angle=0]{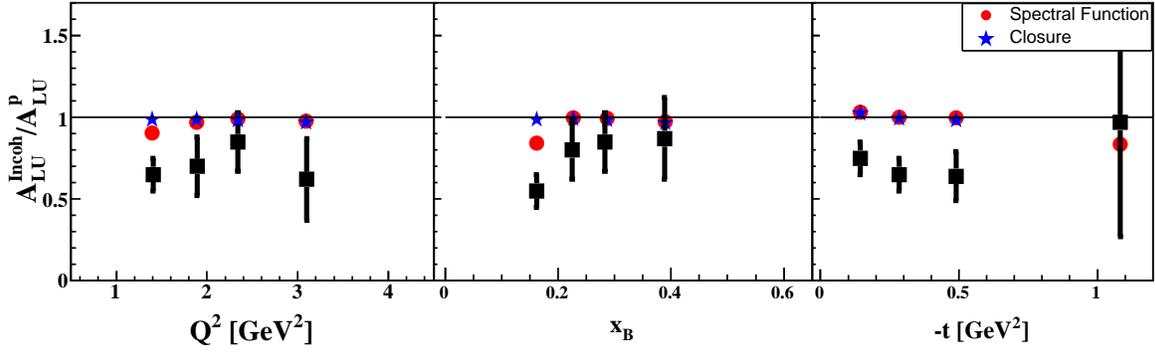}
\caption{
(Color online)
The ratio \eqref{r3}, evaluated using the spectral function (blue triangles) and the closure approximation (black stars) for
$\phi = 90^o$ and using the GK
model for the proton GPD.
From left to right, the quantity is shown in the experimental
$Q^2$, $x_B$ and $t$ bins, respectively.
The results is compared with the same ratio estimated by the EG6
collaboration
(black squares).
\cite{clastobe}.
}
\label{disa}
\end{figure*}

\begin{figure*}[htbp]
\hspace{-.5cm}
\includegraphics[scale=0.85,angle=0]{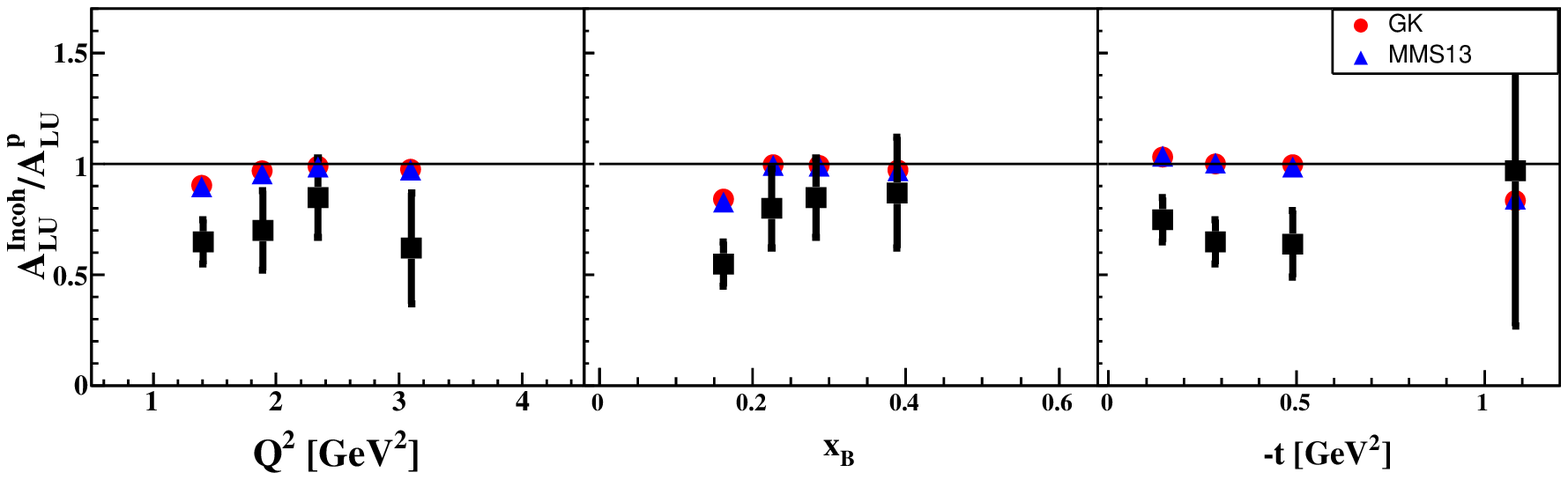}
\caption{
(Color online)
The ratio \eqref{r3}
of the
azimuthal beam-spin asymmetry for the proton in
$^4$He, $A_{LU}^{Inco,h}(K)$, to the corresponding
quantity for the free proton at rest,
for $\phi = 90^o$, using for the proton GPD
the GK model \cite{Goloskokov:2011rd}
(red dots), and the MMS
model
\cite{Mezrag:2013mya}
(blue triangles) compared with the ratio estimated by the EG6
collaboration
(black squares)
\cite{clastobe}.
From left to right, the quantity is shown in the experimental
$Q^2$, $x_B$ and $t$ bins, respectively. 
}
\label{ratioexp}
\end{figure*}
The dots shown in this latter figure are related to
another intriguing observation, obtained
following a procedure used by the experimental collaboration
to expose nuclear effects \cite{Hattawy:2018liu}.
Our BSA for the proton bound in $^4$He
has been divided
by the corresponding quantity for a free proton at rest, using the
GK model, and plotted as a function
of $x_B$.
It is seen that
the results underestimate those
obtained in the analysis of the experimental
collaboration. This points to interesting effects not included in the present IA
scheme, either at the parton level (medium modifications of the parton structure due 
exotic effects, such as
dynamical off-shellness)
or of conventional origin, such as FSI, not yet included
in the calculation.
In Fig. \ref{ratioexp} we show the results obtained with the spectral function and
with either the GK or the MMS model, almost indistinguishable
between themselves.
Clearly, while in the result for $A_{LU}$
the difference between the different models
was in some cases sizable, in this specific
quantity, which can be built in principle
from data taken for protons in $^4$He
and for the free proton at the same kinematics,
this ratio seems to be be essentially
independent on the model used for the nucleon.
In general nuclear
effects are found to be rather small in IA for this quantity,
which seems therefore very promising to
expose exotic nuclear effects.

\begin{figure}[htbp]
\hspace{-.5cm}
\includegraphics[scale=0.45,angle=0]{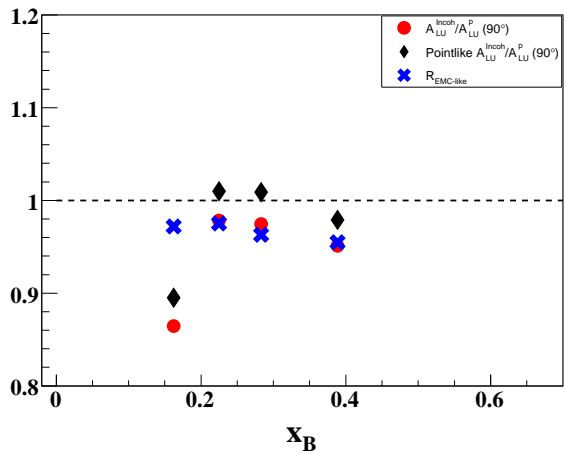} 
\caption{(color online) The ratio 
$A_{LU}^{Incoh} / { A_{LU}^p }$,
Eq. \eqref{aluratio} (red dots),
compared to the result obtained with
pointlike protons (black diamonds)
and to the EMC-like ratio Eq.
\eqref{emclike} (blue crosses).
}
\label{emc}
\end{figure}

To dig further into this interesting result and to realize to what extent
a medium modification of the parton structure is predicted
by our calculation, we observe 
that the ratio \eqref{r3} can be sketched as follows
\begin{equation}
\frac{ A_{LU}^{Incoh} } { A_{LU}^p } 
= \frac{ \mathcal{I}^{^4He} } { \mathcal{I}^{\,p} }
\frac{ T_{BH}^{2\, \, p} } { T_{BH}^{2\, \, ^4He} } \propto \frac
{(nucl.eff.)_{\mathcal{I}}}{(nucl.eff.)_{BH}}\,,
\label{aluratio}
\end{equation}
i.e., it is proportional to the ratio
of the nuclear effects 
on the BH-DVCS interference
to
the nuclear effects on the pure BH cross section.
If the nuclear dynamics modifies $\mathcal{I}$ and the $T_{BH}^2$
in a different way, the effect
can be big even if the parton structure
of the bound proton does not change appreciably.
We analyze this occurrence in 
Fig. \ref{emc}, where, together with the ratio \eqref{r3},
we show two others quantities, as functions of $x_B$.
One of them, labelled "pointlike", is obtained considering in the ratio pointlike protons. It is seen that, at low $x_B$,
where sizable effects are found within our IA
approach, the big effect is still there.
Besides, in the same figure we show an "EMC-like"
quantity, i.e., a ratio of a nuclear parton observable,
the imaginary part of the CFF, 
to the $same$ observable for the free proton:
\begin{equation}
R_{EMC-like}=\frac{1}{{\cal N}} \frac{\int_{exp}
dE \, d\vec{p} \,P^{^4 He} ( \vec{p}, E) \,\Im m \, \mathcal{ H}(\xi',\Delta^2)}{\Im m \, \mathcal{ H}(\xi,\Delta^2)}  \,.
      \label{emclike}
\end{equation}

One should notice that this ratio would be one
if nuclear effects $in$ $the$ $parton$ $structure$ were negligible.
As seen in Fig. \ref{emc}, this ratio
is close to one and it resembles the EMC ratio, for $^4$He, at low $x_B$
(cf the data in Ref. \cite{Seely:2009gt}).
Since in our analysis the inner structure of the bound proton is entirely contained in the CFF and this produces a mild modification,
the sizable effect found for the ratio \eqref{aluratio} for the first $x_B$ bin,
shown in Fig. \ref{emc}, has little to do with the modifications of the parton content driven by the IA and analyzed here. 
Rather, the effect is due to a different dependence on the 4-momentum components, affected by nuclear effects, of the interference and BH terms for the bound proton.

It will be very interesting to study
the ratio \eqref{r3} when consistently collected data will
be available for the proton and for $^4$He,
to look for effects
to be ascribed
to exotic modifications of the parton content or to a
complicated conventional behaviour, beyond IA.


\section{Conclusions}
An impulse approximation analysis,
based on state-of-the-art models for the
proton and nuclear structure, using a conventional description
in terms of nucleon degrees of freedom, has been
thoroughly described.
Recent data on incoherent
DVCS off $^4$He are overall well reproduced. 

The results can be summarized as follows:

i) the main experimental observable, the only one measured so far, the BSA, turns out to be sensitive to
the nucleonic model used, in particular
at low values of $Q^2$; parametrizations
for generalized parton distributions based on high $Q^2$ data
seem to have limited predictive power in the low $Q^2$ sector;

ii) given the present accuracy of the data, the beam spin asymmetry is mildly sensitive
to the details of the nuclear model used
in the calculation, as it can be argued using
a spectral function or its closure approximation.
Results obtained within the spectral function are anyway closer
to a good description of the data;

iii) the behaviour at low $Q^2$ could point also to possible FSI effects, to be investigated, or to other quark and gluon effects. The present accuracy of the data does not allow 
a further analysis towards this direction;

iv) a careful study of nuclear effects in the different
processes contributing to the BSA, the BH in the denominator
and the DVCS-BH-Interference in the numerator, 
has exposed sizable effects; besides, a clear difference is found,
in some kinematical points, if the spectral function
or the closure approximation are used.
The separated measurements of these contributions,
which correspond to those of the differential cross sections
and not only to their ratio, would be very interesting
and deserve to be attempted in the future experiments;

v) all these effetcs actually basically disappear
in the ratio of the interference to the BH contributions.
In our IA approach, the latter ratio represents 
that between the BSA for incoherent DVCS off $^4$He
and coherent DVCS off the free proton.
Its stability against different nuclear and nucleon models,
found in this study,
demonstrates that it can be used to expose interesting exotic
effects beyond the ones included in IA.
We can preliminarly assert that our calculation of this quantity
overestimates the estimate of the experimental collaboration.

We would conlcude that, given the present accuracy of the data,
there is no point in going beyond the exhaustive analysis
presented here.
New tagged measurements with detection of residual nuclear
final states,
planned at JLab \cite{Armstrong:2017zcm} and under study for the future EIC,
will shed more light to this respect.
The presence of specific nuclear final states in these processes
will also make possible a precise evaluation of FSI in terms of few-body
realistic wave functions, allowing for a conclusive comparison with data.

While a benchmark calculation in the kinematics of the
next generation of precise measurements will require
an improved treatment of both the nucleonic and the nuclear
parts of the calculation, such as a realistic evaluation
of the diagonal spectral function of $^4$He,
the straightforward approach proposed here
can be used as a workable framework for the planning of future
measurements. Possible exotic quark and gluon effects in nuclei, not clearly seen within the present 
experimental
accuracy, will be exposed by comparing forthcoming data with our conventional results. To this aim, a novel Montecarlo 
event generator \cite{topeg},
tested so far with our model of the coherent process,
will be used to simulate incoherent DVCS off $^4$He, described
within the approach presented here, to plan the next generation of experiments at JLab and at the future EIC.

\section*{Acknowledgements}
We warmly thank R. Dupr\'e and M. Hattawy for many
helpful discussions and technical information on the EG6 experiment.
S.F. thanks P. Sznajder and C. Mezrag for some tuition on the
use of the virtual access infrastructure 3DPARTONS, funded by the European Union’s Horizon 2020 research and innovation programme under grant agreement No 824093.
This work was supported in part by the STRONG-2020 project of the European Union’s Horizon 2020
research and innovation programme under grant agreement No 824093, Working Package 23, "GPDS-ACT" and by
the project ``Deeply Virtual Compton Scattering off $^4$He", in the programme FRB of the University
of Perugia.

\newpage
\onecolumngrid

\appendix

\section{The convolution formula}

\label{convfor}

Let us start considering the cross section 
$d\sigma^\pm$
appearing in Eq. \eqref{ALU}.
It can be written in a generic frame, for the 
incoherent channel of the DVCS process under scrutiny, namely $  e(k)A(P_A)\rightarrow e(k')N(p_N)\gamma(q_2)X(p_X)  $ off a nuclear target $ A $, in the following way
	\begin{equation}\label{2} 
 (d\sigma^\pm)_{Inc} = (2\pi)^4 
 \frac{1}{4 P_A \cdot k}
 \sum_{\sigma} \sum_{N} \sum_X |{\mathcal{A}}^\pm|^2 \delta ({P}_A+ {k}-{k}' - {p}_X -{p}_N - {q}_2) d{\tilde p}_X d{\tilde k}' d{\tilde q}_2 d{\tilde p}_N 
	\end{equation}
where the dynamical information is encoded in the squared amplitude.
The latter is given by three different contributions, namely $ |{\mathcal A} |^2 =  |{\mathcal A}_{DVCS}|^2+|{\mathcal A}_{BH}|^2+ \mathcal{I}_{BH-DVCS}$.
A generic phase-space integration volume reads
	\begin{equation}\label{lips}
	d{\tilde k} \equiv \frac{d^3k} {(2\pi)^3 2k_0 }.
	\end{equation}
	In Eq. \eqref{2}, the sums are extended to the inner nucleons of type $ N $ in the target, to the polarization $\sigma$ of the final detected proton and to the undetected nuclear system $X$. The status $f$ of the latter is identified by a set $\{\alpha_f\}$ of discrete quantum numbers and by the excitation energy $E_f$, for which discrete and continuous values are possible. One has therefore, in Eq. \eqref{2},
	\begin{equation}\label{somma_am1}
	\sum_X d{\tilde p}_X \rightarrow \sum_{f}  \displaystyle\sum_{\{\alpha\}_f} \sumint_{E_f}\rho(E_f)\, d{ \tilde p}_{f} \, ,
	\end{equation}
	where $\rho(E_f)$ is the density of final states.
The amplitudes $ \mathcal{A}_{BH} $ and $ \mathcal{A}_{DVCS} $ 
appearing in Eq.\eqref{2} are given by the contraction of a leptonic tensor ($ L^\nu_{DVCS} / Q^2$ and $ L^{\mu \nu}_{BH}/\Delta^2 $
 for $DVCS$ and $BH$, respectively) with the appropriate hadronic tensor. For a generic DVCS process of a target $ A $ with initial(final) polarization $ S(S') $ reads
	\begin{equation}  
	T_{\mu\nu}^{DVCS}(P_A,\Delta,q, S,S')=\int dr e^{iq\cdot r} \langle P'_A S'| T \{	\hat{J}_\mu(r) \hat{J}_\nu (0) \}|P_A S\rangle \, .
	\end{equation}
Since a convolution formula with the same structure can be obtained for
any of the $DVCS$, $BH$ and interference terms exploiting the same steps, 
to fix the ideas in what follows we specify our treatment to the $DVCS$
part.	
Let us consider therefore the
scattering amplitude of the incoherent DVCS process off an $ ^4 $He target, i.e $ e(k)^4He(P_A)\rightarrow e(k')N(p_N)\gamma(q_2)X(p_X) $
	\begin{equation}\label{DVCSamplitude}
	{\mathcal A}_{DVCS}^{A,N,f} = -ie \sum_{\lambda'}\bar{u}(k',\lambda')\gamma_\mu u(k,\lambda)\frac{1}{Q^2}T_{A,N,f}^{\mu \nu}\epsilon_\nu^*(q_2) =\frac{\epsilon_\nu^*(q_2)}{Q^2} L_{\mu}^{DVCS} (\lambda)T_{A,N,f} ^{\mu \nu}
\, ,
	\end{equation}
	where it appears the hadronic tensor $T ^{A,N} _{\mu \nu} $, defined in terms of  
	\begin{equation}\label{hadronic}
	T^{A,N} _{\mu \nu} = \int d^4r e^{-iq \cdot r } H_{\mu\nu}^{A,N}
	\, ,
	\end{equation}
	being $ H_{\mu\nu}^{A,N}  $ the matrix element of $ \hat{O}^N = T\{\hat{J}^N_\mu(x)\hat{J}^N_\nu(0)\} $ properly evaluated between the states describing the initial and the final nucleon $N$ in the nucleus $A$, respectively. Here and in the following, we are assuming that the interaction goes through the nucleons in the nucleus, which are the only degrees of freedom in the present
	Impulse Approximation (IA).
	Disregarding for the moment the integration on $ x $, let us focus on the matrix element $ H_{\mu\nu} ^{A,N,f} $. \\
We will use in the following the standard covariant normalization of the states
\begin{equation}\label{7}
\langle {p} \sigma| {p}' \sigma' \rangle 
= (2\pi)^3 2p_0 \delta(\vec{p}' - \vec{p} ) \delta_{\sigma,\sigma'}
\end{equation} 
and the notation
$  
\sum_{{p}} = \int d \tilde p
$ is used.
The matrix element in Eq. \eqref{hadronic} is therefore 
	\begin{equation}\label{4}
	H^{A,N,f}_{\mu\nu} = \langle {p}_N\sigma, \, {p}_f {\{ \alpha_f\} }E_f|\hat{O}_N|{P}_A \rangle\, ,
	\end{equation}
	where the final state contains the detected nucleon with momentum $ p_N $ and polarization $ \sigma $ and the $ A-1 $- body system described by a set of quantum numbers $ \{ \alpha_f \} $, whose constituents are moving with momenta $ p_f $. 
Let us insert to the left and to the right-hand sides of the hadronic operator two complete sets of states; the first set corresponds to the nucleon $N$, supposed free, interacting with the virtual photon, whose completeness reads
	\begin{equation} \label{5}
	\sum_{{p}'_N \sigma'}| {p}'_N \sigma'\rangle \langle {p}'_N\sigma'|= 1,
	\end{equation}
	 while the completeness of the second set of states, describing the hadronic undetected system, is given by:
\begin{equation}  
\sum_{\{\alpha_f\}} \sumint_{E_f} \rho(E_f)\sum_{{p}_f} | {{p}_f} \{ \alpha_f \}  {E_f}  \rangle  \langle {{p}_f} \{ \alpha_f \}  {E_f}|=1~.
\end{equation}
	Now let us use the IA. This means that the interaction goes only through the nucleons, as already said, and that the final state can be written as a  tensor product 
	\begin{equation}\label{ia}
	|{p}_N \sigma, {p}_f \{\alpha_f\} E_f \rangle = |{p}_N \sigma \rangle  \otimes |{p}_f \{\alpha_f \} E_f \rangle ,
	\end{equation}
	i.e., the interactions between the particles in the final state (FSI) have been neglected.
	At the light of these facts, we arrive to the following formula
	\begin{eqnarray} 
	H^{A,N,f}_{\mu\nu}&=& \sum_{\{\alpha'_f\}} \sumint_{E'_f} \rho(E'_f)\sum_{{p'}_{f}} \sum_{{p}'_N \sigma '}\langle {p}_N\,  \sigma | \langle {p}_f \{{\alpha_f}\} E_f|
	\hat{O}_N | {p'}_f \{\alpha_f '\} E_f '\rangle |{p'}_N \sigma' \rangle \langle {p'}_N \sigma'| \langle {p'}_f 
	\{\alpha'_f\} E_f '|{P}_A \rangle.
	\end{eqnarray} 
	Now, assuming in IA that the one-body operator $\hat{O}_N $ acts only on the nucleonic states, we can consider the normalization \eqref{7} to perform trivially some integrals,
	obtaining the following form:
	\begin{eqnarray}\label{18}
	H^{A,N}_{\mu\nu} &=& \sum_{{p}_N^{'} \sigma'} \langle {p}_N \sigma| \hat{O}_N|{p}_N'\sigma'\rangle \langle {p}_f \{\alpha_f\} E_f|\langle {p}_N' \sigma'|{P}_A \rangle. 
	\end{eqnarray}
A relevant issue has to be discussed at this point. 
Since relativistic nuclear wave functions for three and four body systems are not at hand,
in the following we will be forced to use non relativistic wave functions
in the overlaps of the above equation. Therefore, we will use for the states in the overlap a non relativistic normalization
\begin{eqnarray}
\langle \vec p s | \vec p' s' \rangle = \delta( \vec p - \vec p') \delta_{ss'} \, .
\label{nr}
\end{eqnarray}
For the same reason, in the overlap we can disentangle the global motion from the intrinsic one
\begin{equation}\label{intrinsic}
|{p}_{f} \{\alpha_f\} E_f \rangle = | \Phi^{ \{ \alpha_f \}}_{E_f} ( {p}_{f'} , \sigma_{f'} );\, {p}_x s_x\rangle \, ,
\end{equation} 
where $ \Phi^{ \{ \alpha_f \}}_{E_f} $ represents the intrinsic  motion of the final system, described by $ A-1 $ fully interacting particles, with $ A-2 $ independent
momenta $ {p}_{f'}$ and intrinsic quantum
numbers $\sigma_{f'}$, while $ {p}_x $ and
$s_x$ specify the state of the center of mass 
of the $ A-1 $-body system
	(for an easy notation, in the following, we will denote the intrinsic wave function simply with the ket $ | \Phi^{ \{ \alpha_f \}}_{E_f} \rangle $ instead of $ | \Phi^{ \{ \alpha_f \}}_{E_f} ( {p}_{f'}, \sigma_{f'}) \rangle $).
	 In this way the overlap becomes 
	\begin{equation}\label{overlap}
\{ \langle \Phi^{ \{ \alpha_f \}}_{E_f}\,  {p}_x\, s_x| \} \langle \vec {p}_N\,' \sigma'|\vec {P}_A \rangle =
[(2 \pi)^{3/2}]^4 \sqrt{2M_A}\sqrt{2p'^0_N}\sqrt{2p_x^0}
\sqrt{2p_f^0}
\langle \vec{p}_N' \sigma', \Phi^{ \{ \alpha_f \}}_{E_f} |{\Phi}_A \rangle\, \delta (\vec{P}_A- \vec{p}_x -\vec{p}_N' )\delta_{\sigma', \, -\sigma_f - s_x}\, ,
\end{equation}
	where the momentum delta function accounts for the center of mass free motion and 
	$ \Phi_A $ is  the intrinsic wave function of the target nucleus.
	The other delta function yields a formal condition to be fulfilled between the discrete
	quantum numbers appearing in the overlap. The terms at the beginning of the
	r.h.s. account for the chosen non relativistic normalization of the states Eq. \eqref{nr}.
	In this way, from Eq. \eqref{18} we get 
	\begin{eqnarray}\label{20}
	H_{\mu\nu}^{A,N,f} &=& 
	\sum_{ \sigma'}\sum_{{p}_N'} 
	[(2 \pi)^{3/2}]^4 \sqrt{2M_A}\sqrt{2p'^0_N}\sqrt{2p_x^0}
\sqrt{2p_f^0}
\langle {p}_N \, \sigma|\hat{O}_N| {p}_N' \sigma' \rangle
\langle \vec{p}_N' \sigma', \Phi^{ \{ \alpha_f \}}_{E_f} |{\Phi}_A \rangle\, \delta (\vec{P}_A- \vec{p}_x -\vec{p}_N' )\delta_{\sigma', \, -\sigma_f - s_x}\, ,
\nonumber
	\end{eqnarray}
	so that the complete expression for the hadronic tensor in the incoherent DVCS channel becomes:
	  \begin{eqnarray}  
		T^{A,N}_{\mu\nu} &=& \sum_{\sigma'} \sum_{{p}_N'}
				\int dr e^{ iq\cdot r} 
		[(2 \pi)^{3/2}]^4 \sqrt{2M_A}\sqrt{2p'^0_N}\sqrt{2p_x^0}
		\nonumber
\sqrt{2p_f^0}
\langle {p}_N \, \sigma|\hat{O}_N| {p}_N' \sigma' \rangle
\langle \vec{p}_N' \sigma', \Phi^{ \{ \alpha_f \}}_{E_f} |{\Phi}_A \rangle\, \delta (\vec{P}_A- \vec{p}_x -\vec{p}_N' )\delta_{\sigma', \, -\sigma_f - s_x}\, ,			
	 \end{eqnarray}
which can be inserted in the DVCS amplitude Eq. \eqref{DVCSamplitude} obtaining
	\begin{eqnarray}\label{45}
	{\mathcal A}^{A,N,f,\lambda}_{DVCS} 
	& = & -ie \sum_{\lambda'}\frac{\bar{u}(k',\lambda')\gamma_\mu u(k, \lambda) }{Q^2} \sum_{\sigma'}\int dr e^{iq\cdot r}\sum_{ {p}'_N}\langle {p}_N \, \sigma|T ({\hat{J}_N^\mu(r)\hat{J}_N^\nu(0)})|{p}_N' \sigma' \rangle  \\ 
	& \times& 
	[(2 \pi)^{3/2}]^4 \sqrt{2M_A}\sqrt{2p'^0_N}\sqrt{2p_x^0}
\sqrt{2p_f^0}
\langle {p}_N \, \sigma|\hat{O}_N| {p}_N' \sigma' \rangle
\langle \vec{p}_N' \sigma', \Phi^{ \{ \alpha_f \}}_{E_f} |{\Phi}_A \rangle\, \delta (\vec{P}_A- \vec{p}_x -\vec{p}_N' )\delta_{\sigma', \, -\sigma_f - s_x}
		\epsilon^*_\nu \, .
		\nonumber
	\end{eqnarray}
		
		Now, let us consider the squared amplitude appearing in the expression of the cross section, Eq. \eqref{2}
	\begin{eqnarray}\label{squared}
	|{\mathcal A}^{A,N,f,\lambda}_{DVCS} 
	|^2 & = &  
	(2 \pi)^{12} 2M_A 
	\sum_{ \sigma'' } \sum_{ \sigma'}  \sum_{  {p}_N' } \sum_{  {p}_N''} 
	2p_x^0 2 p_f^0 \sqrt{2 p_N'^0} \sqrt{2 p_N^{''0}} 
	\bigg|\mathcal{A}_{DVCS}^{N, \lambda}  ({p}_N, {p}'_N,\sigma, \sigma')\bigg|^2\langle \vec{p}_N' \sigma', \Phi^{ \{ \alpha_f \}}_{E_f}  |\,{\Phi}_A\rangle  \\ 
	& \times & \langle {\Phi}_A | \,\vec{p}_N'' \sigma'', \Phi^{ \{ \alpha_f \}}_{E_f}  \rangle  \delta ({{\vec P}}_A- \vec{p}_x -{\vec p}_N' ) \delta({{\vec P}}_A- {\vec p}_x -
	{\vec p}_N'' ) \delta_{\sigma', \, (-\sigma_f-s_x)} \delta_{\sigma'', \, (-\sigma_f-s_x)}
	\, , \nonumber
	\end{eqnarray}
	where the squared DVCS amplitude off a nucleon is given by 	\begin{eqnarray}\label{nucleon}
	|{\mathcal A}_{DVCS}^{N\,\lambda} ({p}_N,{p}_N',\sigma') |^2  & = & \sum_{\sigma} |{\mathcal A}_{DVCS}^{N\,\lambda} ({p}_N,{p}_N',\sigma, \sigma') |^2  \\& = &- \frac{g^{\mu \nu}} {Q^4} \sum_{\sigma} 
	\int dr' e^{-iq\cdot r'}  \int dr e^{iq\cdot r} L_{DVCS}^\rho (\lambda) L_{DVCS} (\lambda)^{\sigma \dagger}
	 \langle  {p}_N  \sigma |\hat{O}^N_{\mu \nu}| {p} ' _N\sigma '\rangle \langle{p}_N'  \sigma ' |
	\hat{O}^{N \dagger}_{\rho \sigma}| {p}_N \sigma \rangle \, .
	\nonumber
	\end{eqnarray}
	In this way, substituting the obtained expression in the cross section \eqref{2}, taking into account that,
	due to the separation of the global motion from the intrinsic one in the $A-1$ system 
the sum \eqref{somma_am1} reads:

		\begin{equation}  
	\sum_X d{p}_X \rightarrow \sum_x \sum_{f'} \displaystyle\sum_{\{\alpha\}_f} \sumint_{E_f}\rho(E_f) d{\tilde p}_x d{ \tilde p}_{f'} \, ,
	\end{equation}
	
	and using the delta functions we arrive to
	\begin{eqnarray}  \label{dsig}
(d\sigma^A)_{Inc} &=&  (2\pi)^4 
\frac{1}{4 P_A \cdot k}
\sum_N \sum_{\sigma'} \sum_x \sum_{f'}  d \vec{p}_x  d \vec{p}_{f'} \sum_{\{\alpha_f\}} \displaystyle\sumint_{E_f} \rho(E_f)  \bigg|\mathcal{A}_{DVCS}^{N,\lambda} 
({p}_N,\, {p_N}', \, \sigma' ) \bigg|^2 
\frac{M_A}{p_N^{' 0}}
\\ & \times&   \langle \vec{p}_N' \sigma', \Phi^{ \{ \alpha_f \}}_{E_f} |{\Phi}_A \rangle  
	\langle {\Phi}_A| \vec{p}'_N\sigma', \Phi^{ \{ \alpha_f \}}_{E_f} \rangle  \delta ({P}_A+ {k}-{k}' - {p}_X -{p}_N - {q}_2) d{\tilde k}' d{\tilde q}_2 d{\tilde p}_N \, ,
	\nonumber
		\end{eqnarray}
		where one has to read $ \sigma' \equiv-( \sigma_f+s_x )$.	Finally, defining the diagonal spectral function as
	\begin{equation}  
	P^{^4He}_N(\vec{p}_N,E)= 
	\sum_{  \{ \alpha_f \} } 
	\int d \vec p_{f'}
\rho(E) \langle \vec{p}'_N \sigma', \Phi^{ \{ \alpha_f \}}_{E} |{\Phi}_A \rangle  
	\langle {\Phi}_A| \vec{p}_N' \sigma', \Phi^{ \{ \alpha_f \}}_{E} \rangle ~,
	\end{equation} 
	where the standard removal energy definition $E\equiv E_f=
|E_A| - |E_{A-1}| + E_f^*$
	has been adopted,
	 the cross section \eqref{dsig} can be rewritten in the following compact way
\begin{eqnarray}
d\sigma^{\lambda}_{Inc} &=& \frac{1}{4 P_A \cdot k} \sum_{\sigma '} \sum_N
	\sumint_{E} \int d{\vec p}P_N^{^4He}(\vec {p},E)
	\frac{M_A}{p_0}
	|A_{DVCS}^{N ,\lambda}({p}, {p}_N,\sigma')|^2 (2\pi)^4\delta({P}_A+{q}-{p}_N-q_2-p_X)d{\tilde k}'d{\tilde p_N} d{\tilde q}_2
	\nonumber \\
 &=&
 \frac{1}{4 P_A \cdot k}
 \sum_{\sigma'} \sum_N 
 \sumint_{E}
 \int d{\vec p}P_N^{^4He}(\vec{p},E)\frac{M_A}{p_0}|A_{DVCS}^{N, \lambda}({p}, {p}_N,\sigma')|^2 (2\pi)^4\delta({p}+{q}-{p}_N-q_2)d{\tilde k}'d{\tilde p}_N d{\tilde q}_2
 \label{convolution}
\end{eqnarray}
where we used that $ \vec p_X = \vec p_f+ \vec p_x $ and that $ \vec p_f = \sum_{f'} \vec{p}_{f'} =0 $. Besides, we also made use of the condition given by \eqref{overlap}, i.e  $  \vec{p}_N= \vec{P}_A -\vec{p}_x  $; in addition to this,
in the spirit of the IA, we have energy conservation at the
nuclear vertex, so that $ {p}^0_N= {P}_A^0 -
{p}_x^0  $. In the last step we changed the name of the
integration variables defining a four momentum of an off-shell
nucleon, $p=(p_0,\vec p)$.


Now, keeping in mind that for a coherent DVCS process off a single nucleon the analogous cross section reads
\begin{equation} \label{nucmov}
    d\sigma_{Coh}^{\lambda,N} = \frac{1}{4 p \cdot k} |A_{DVCS}^{N, \lambda}({p}, {p}_N,\sigma')|^2 (2\pi)^4\delta({p}+{q}-{p}_N-q_2)d{\tilde k}'d{\tilde p}_N d{\tilde q}_2
\end{equation} 
we can rewrite Eq. \eqref{convolution} as a clear convolution formula between the spectral function $P_N^{^4He}$ of the inner nucleons and the cross section for a DVCS process off an off-shell nucleon, namely
\begin{eqnarray}\label{conv}
d\sigma^\lambda_{Inc} &=&  \sum_\sigma \sum_N 
 \sumint_{E}
 \int d{\vec p} \frac{p\cdot k}{P_A\cdot k} \frac{M_A}{p_0} P_N^{^4He}(\vec{p},E) \, d\sigma_{Coh}^{\lambda, N} \, .
\end{eqnarray}
If the above equation is evaluated in the target rest frame, it becomes
\begin{eqnarray}
d\sigma^\lambda _{Inc} &=&  \sum_\sigma \sum_N 
 \sumint_{E}
 \int d{\vec p} \, \frac{p\cdot k}{p_0E_k} \, P_N^{^4He}(\vec{p},E) \, d\sigma_{Coh}^{\lambda, N}
\, .
\end{eqnarray}


We have now to obtain a workable expression for the differential 
cross section
to be used in the actual calculation and to be related to experimental data for the beam spin asymmetry. To this aim,
let us rewrite the invariant phase space 
($LIPS$) for the coherent cross section for a moving nucleon, Eq.
\eqref{nucmov}, that reads explicitly
\begin{equation}  
LIPS= d \tilde k' d \tilde p_N d \tilde q_2= 
\frac{d^3k'}{2E'(2\pi)^3}
\frac{d^3p_N}{2E_2(2\pi)^3} 
\frac{d^3q_2}{2\nu' (2\pi)^3} \, .
\end{equation}
Let us choose, as everywhere in this paper, the target rest frame where the spacelike virtual photon propagates along the negative z-axis,i.e
$  
q_1=(k-k')= (\nu, 0,0,-q_1^z)$ with $ Q^2 = -q_1^2 \, . 
$
In this frame, the kinematical variables are
(it is assumed that $\vec k$ lies in the $xz$ plane):
\begin{eqnarray}  
k &=& (E_k,E_k \sin \theta_e,0,E_k\cos\theta_e)\\
k' &=& (E',\vec{k}')\\
P_A &=& (M_A, \vec{0})\\
p_N & =& (E_2, |\vec{p}_N|
\sin \theta_N \cos \phi_N, |\vec{p}_N|\sin \theta_N \sin\phi_N,|\vec{p}_N|\cos\theta_N )
\label{a31}
\\
q_2 &= & (\nu', \vec{q}_2)
\end{eqnarray}
We have to specify the components of the 4-momentum of the bound nucleon.
In this framework, the energy conservation in the electromagnetic
nuclear vertex yields
\begin{equation}  
p_0= M_A-p^{0}_x = M_A -\sqrt{M_{A-1}^{*2}+\vec{p}_{A-1}^2 }\approx M-E- K_R \, .
\end{equation}
The interacting  nucleon has 3-momentum $ \vec{p} $ ( $ \vartheta $ is the polar angle of $ \vec{p}$, so that the angle between $ \vec{p} $ and $ \vec{q} $ is $ \pi -\vartheta $ )  and       $ K_R $ is the kinetic energy of the recoiling $ A-1 $ body system.
The experimental cross section is 4 times differential
in the variables $x_B=Q^2/(2 M \nu)$, $\Delta^2= (q-q_2)^2$,$\phi_N$, $Q^2$. 
In addition to these variables, in the following we will make use
of the quantity: $\epsilon=2Mx_B/Q$.
The LIPS, in terms of these variables,
read

\begin{eqnarray} \label{newlips}
LIPS =  J(p_N \rightarrow \Delta^2) d \Delta^2 d \cos \theta_N d \phi_N  \frac{Q^2}{2(2\pi)^3 \,  { 2}M \,{2} E_k x_B^2} d Q^2 d x_B d \phi_{k'}
\frac{d^3q_2}{2\nu' (2\pi)^3} \, ,
\end{eqnarray}

where the term $J(p_N \rightarrow \Delta^2)$ is proportional to the jacobean of the transformation and reads,
since the process takes place on a moving nucleon,

\begin{eqnarray}\label{jaco}
J(p_N \rightarrow \Delta^2) = \frac{1}{4 ( 2 \pi)^3} {\bigg|} \frac{|\vec{p}_N|^2}{ |\vec{p}|\cos\theta_{\hat{pp_N}}
E_2-p_0 |\vec{p}_N|  } {\bigg|} \, ,
\end{eqnarray}

where

\begin{equation}\label{cospp'}
\cos \theta_{\hat pp_N}= \cos \theta_N \cos\vartheta + \sin \theta_N\sin\vartheta \cos(\phi_N-\varphi) \, .
\end{equation}

Substituting Eq. \eqref{newlips} in Eq. \eqref{convolution},
using the delta function on the three-momenta
to obtain $\vec q_2= \vec p + \vec q - \vec p_N$,
and using this result in the delta function on the energy
variables to integrate on $\cos \theta_N$, one finally
obtains the cross section in the nuclear rest frame

\begin{eqnarray}\label{diff-inc}
\frac{d\sigma^\lambda_{Inc} }{d{x}_BdQ^2 d{\Delta}^2 d\phi_N}
&=&
{ \frac{Q^2}{32 E_k^2 M ( 2 \pi)^2 x_B^2}}
\sum_\sigma \sum_N
	\sumint_{E} \int_{exp} d{\vec p}P_N^{^4He}(\vec {p},E)
	\nonumber\\
	& \times &
	|A_{DVCS}^{N ,\lambda}({p}, {p}_N,\sigma)|^2
{\cal G}( {p},|\vec{p}_N|,
K) \, .
\end{eqnarray}

In the equation above, we have defined
the set of kinematical variables $K=\{x_B=Q^2/(2 M \nu),Q^2,t,\phi\}$ and
\begin{eqnarray}\label{gtheta}
{\cal G}( p,|\vec{p}_N|,
K) \, 
&=&  \frac{1}{p_0}
\int(2\pi)^4 \delta^4 (p-p_N-q_2+q)
J(p_N \rightarrow \Delta^2) 
\frac{d^3q_2}{2(2\pi)^3\nu'} d\cos{\theta}_N  
\nonumber\\ &=&
\pi \bigg |\frac{1}{|\vec p_N|(|\vec p|(\sin\vartheta\cot\bar{\theta}_N\cos
(\phi_N-\varphi )- \cos\vartheta)-2q_1^z)} \bigg| J(\cos{\bar{\theta}}_N) \, ,
\end{eqnarray}
where $ J(\cos \bar{\theta}_N)$
is the expression $J(p_N \rightarrow \Delta^2) $
evaluated for $\cos {\bar{\theta}}_N $, which is obtained from 
the energy conservation condition
\begin{eqnarray}\label{zeros}
\sqrt{|\vec p|^2+|\vec p_N|^2+|q_1^z|^2 -2 |\vec p||\vec p_N|\cos\theta_{pp_N}-2|\vec p_N|q^z \cos\theta_N +2 |\vec  p|q_1^z \cos\vartheta}-p_0+E_2-\nu=0 \, ,
\end{eqnarray}
where Eq. \eqref{cospp'} is exploited.
We note that the quantity $|\vec p_N|$ can be obtained from the relation
\begin{eqnarray}\label{pN}
{\Delta}^2 = (p_N-p)^2=
M^2 +p_0^2 - |\vec{p}|^2 -2p_0 \sqrt{ M^2+|\vec p_N}|^2  + 2|\vec{p}_N| |\vec{p}| \cos\theta_{\hat{pp_N}} \, ,
\end{eqnarray}
where the expression for the angle between $\vec p $ and $\vec{p}_N$ 
is given by Eq. \eqref{cospp'}.
The values of $\cos \bar{\theta}_N$  and $|\vec p_N|$ to be considered
in the following are obtained through the numerical solution
of the system of equations
\eqref{zeros} and \eqref{pN}.

In order to have a clear comparison between our cross section and that for a DVCS process off a proton at rest, i.e.
\begin{equation}
 \frac{d\sigma^\lambda_{\text rest} }{d{x}_B dQ^2 d{\Delta}^2 d\phi_N}
=
\frac{ \alpha ^3 x_B y^2}{8 \pi Q^4 \sqrt{1 + \epsilon^2}} 
 \bigg|\frac{\mathcal{A}_{DVCS}}{e^3} \bigg|^2 \, ,  
\end{equation}
let us rewrite Eq. \eqref{diff-inc} in the following way, corresponding to
Eq. \eqref{crossl}

\begin{eqnarray}\label{diff-inc-1}
\frac{d\sigma^\lambda_{Inc} }{d{x}_BdQ^2 d{\Delta}^2 d\phi_N}
&=&
\sum_N
	\sumint_{E} \int_{exp} d{\vec p} \, P_N^{^4He}(\vec {p},E)	\bigg |{A_{DVCS}^{N ,\lambda}({p}, {p}_N, K)}\bigg|^2
g(E,\vec{p},K) \, ,
\end{eqnarray}
where 
\begin{eqnarray}
g(E,\vec{p},K) =
{\frac{ \alpha ^3 Q^2 \pi }{ 2E_k^2 M x_B^2 e^6} }
\, {\cal G}( {p},|\vec{p}_N|,
K) 
\end{eqnarray}
and the sum over the proton polarization 
in Eq. \eqref{diff-inc}
has been absorbed by the squared amplitude.
The label $exp$ in the above equation describes the fact that the integration
region is restricted to the components of $\vec p$ and to
the values of E fulfilling
the conditions \eqref{zeros} and \eqref{pN}.

\section{Scattering amplitudes for the proton
bound in $^4$He}

\label{scattamp}

In this appendix we report the expression to be used for the amplitudes relevant to photon-electroproduction 
off a bound off-shell proton in $^4$He.
This will be achieved generalizing the result obtained 
for  a free proton at rest.
Let us recall first the main
formalism for that case.

\subsection{Formalism for the proton in the rest frame.}

Let us study coherent DVCS (e + p  $ \rightarrow$ e'+$\gamma$+p') off a proton at rest, with
4-momentum $p_1 = (M, \vec{0})$.
Using the notation and the reference frame discussed in the text and in the previous appendix,
the general cross section,

 \begin{equation}\label{crosscoh}
 d \sigma= \frac{1}{4 p_1 \cdot k}|{\cal T}|^2\frac{d^3k'}{2 E' ( 2 \pi)^3} \frac{d^3 p_N}{2 E_2 (2 \pi)^2} \frac{d^3 q_2}{2 \nu ' ( 2 \pi ^3)} \delta^4(p_1 + k-k'-p_N -q_2) \, ,
 \end{equation} 
 with $ |{\cal T}|^2 = {\cal T}_{BH}^2+{\cal T}_{DVCS}^2+\mathcal{I}_{BH-DVCS} $.
 Here and in the following, if not differently stated, we take into account terms of order	$ \frac{\Delta^2}{Q^2},
 \,\, \epsilon^2\,\,$ with $\epsilon=\frac{2 M x_B}{Q} $, so that
 the virtual photon and the final photon have 4-momentum components
 
 	\begin{eqnarray}
	q_1 & = & \bigg( \frac{Q}{\epsilon},0,0, -\sqrt{1+\epsilon^2} \frac{Q}{\epsilon}\bigg) \, , \nonumber \\
	q_2 & = & \bigg(\frac{Q}{\epsilon} +\frac{\Delta^2}{2 M}\bigg)\bigg(1, -\sin(\theta_\gamma)\cos(\phi_N),-\sin(\theta_\gamma)\sin(\phi_N) , \cos(\theta_\gamma)\bigg) \, ,
	\end{eqnarray}
respectively,
and the struck proton has final momentum \eqref{a31} 
	with
	$|\vec p_N| = \sqrt{-\Delta^2 \bigg( 1- \frac{\Delta^2}{4 M^2} \bigg)}$, $\cos \theta_N = -\frac{\epsilon^2Q^2(1 - \Delta^2/Q^2) -2 x_B\Delta^2}{4 x_B M |\vec{p}_N|\sqrt{1+\epsilon^2}}$.
	We note that the electron scattering angle is given by $\cos \theta_e =- \frac{1+y \epsilon^2/2}{\sqrt{1+\epsilon^2}}$, and
we remind that
	 $P = p_1 + p_N$,
	 $q = \frac{q_1+q_2}{2}$.

	In the following, we will review the computation of the BH and Interference amplitudes for the proton at rest, and their decomposition in Fourier harmonics depending on $ \phi_N $, which turns out to be equal to $\phi$ in our framework.
In the following section of the Appendix, we will generalize these expressions to describe a moving, bound proton.
We do not treat the pure DVCS process because it is expected to be very small in the JLab kinematics of interest here and it has been neglected in our analysis.
	
	\subsubsection{Bethe-Heitler term}

The amplitude corresponding to the diagrams in Fig. \ref{dvcshb}  can be computed exactly starting from
	\begin{equation}\label{tbh0}
	\mathcal{T}_{BH} = \frac{e^3}{\Delta^2}  \epsilon^{* \mu}(q_2) \bar{u}(k',s') \bigg( \gamma_\mu \frac{1}{\slashed{k} -\slashed{\Delta}} \gamma_\nu +\gamma_\nu \frac{1}{\slashed{k}' +\slashed{\Delta}} \gamma_\mu\bigg)u(k,s)
	{\cal J}^\nu.
	\end{equation}

	The $ \phi $ dependence of the amplitude comes from the lepton propagators  (cf. Fig. \ref{dvcshb}) which read:
	\begin{eqnarray}\label{propbh}
	P_1(\phi) & = & \frac{(k'+\Delta)^2 }{Q^2}= 1 + \frac{ 2 k\cdot \Delta }{Q^2}= - \frac{1}{y(1 + \epsilon^2)} ( J + 2 \tilde K \cos(\phi)) \, , \\
	P_2(\phi) &=& (k-\Delta)^2/Q^2 = \frac{\Delta^2 -2 k\cdot \Delta }{Q^2}=1+\frac{\Delta^2}{Q^2} + \frac{1}{y(1 + \epsilon^2)} ( J + 2 \tilde K \cos(\phi)) \, ,
	\end{eqnarray}
	where we have rewritten the scalar product $ k \cdot \Delta $ in terms of the following quantities: 
	\begin{eqnarray}\label{K}
	 J  = & \bigg(1- y - \frac{y \epsilon^2}{2}\bigg)\bigg( 1+ \frac{\Delta^2}{Q^2}\bigg) - (1 - x) (2-y) \frac{\Delta^2}{Q^2} \, ,\\ 
	 \tilde K^2= & - \frac{\Delta^2}{Q^2} (1-x) (1 - y - \frac{y^2 \epsilon^2}{4})\bigg( 1-\frac{Q^2}{\Delta^2} \frac{2(1-x_B)(1-\sqrt{1+\epsilon^2}) +\epsilon^2}{4 x_B(1-x_B)+\epsilon^2}\bigg)\\ &\bigg[\sqrt{1+\epsilon^2} +\frac{4x_B(1-x_B)+\epsilon^3}{4(1-x_B)}\bigg( \frac{\Delta^2}{Q^2}-\frac{2(1-x_B)(1-\sqrt{1+\epsilon^2}) +\epsilon^2}{4 x_B(1-x_B)+\epsilon^2})\bigg)\bigg]\,.
	\end{eqnarray}
	Ignoring the electron mass, Eq. \eqref{tbh0} yields:

	\begin{equation}\label{squaredbh}
	    |\mathcal{T}_{BH}|^2 =  \frac{e^6}{\Delta^4}  \sum_{s',S'}(-g^{\mu \mu'}) L_{\mu \nu}^{ \dagger} L_{\mu' \nu'}^{ } {\cal J}_{}^{\nu}
	    {\cal J}_{}^{\dagger \nu'} = \frac{e^6}{\Delta^4} \mathcal{J}^{BH}_{\nu' \nu}   \mathcal{L}^{\nu \nu'}_{BH} \, ,
	\end{equation}
	where, in the last step, the hadronic and the leptonic tensors obtained summing over the final proton and electron polarizations, 
	$S'$ ans $s'$,
	respectively, read
	\begin{align}\label{adrorest}
	   \mathcal{J}_{BH}^{\mu \nu} = & \frac{1}{2} \bigg [F_1(\Delta^2)^2+(F_1(\Delta^2)+F_2(\Delta^2))^2-\frac{\Delta ^2 }{4 M^2} F_2(\Delta^2)\bigg] (p_1^{\nu }
   p_N^{\mu }+p_1^{\mu } p_N^{\nu })+\frac{\Delta ^2}{2}  (F_1(\Delta^2)+F_2(\Delta^2))^2 g^{\mu \nu }+ \nonumber \\ 
   & \frac{1}{2}
   \bigg(F_1(\Delta^2)^2-(F_1(\Delta^2)+F_2(\Delta^2))^2-\frac{\Delta ^2 }{4 M^2}F_2(\Delta^2)\bigg) (p_1^{\mu } p_1^{\nu
   }+p_N^{\mu } p_N^{\nu }) \, ,
	\end{align}
where	 $ F_1$ and $F_2 $ are the nucleonic Dirac and Pauli form factors, and
	\begin{align}\label{leptobh}
	    \mathcal{L}^{\mu \nu}_{BH} &= \frac{8}{Q^4 P_1(\phi)P_2 (\phi)} \bigg\{  \left [ 2  {k}\cdot  {\Delta } +Q^2\bigg(1-\frac{\Delta^2}{Q^2}\bigg)  \right ] 
	    (       {k'}^{\nu }        {q_2}^{\mu }+       {k'}^{\mu }{q_2}^{\nu } )- Q^2\bigg(1-\frac{\Delta^2}{Q^2} \bigg) (   {k}^{\nu } {k'}^{\mu }+       {k}^{\mu }        {k'}^{\nu } )+
	    \nonumber
	    \\& 
	    -2 {g}^{\mu \nu }
        ( ( {k'}\cdot{q_2} )^2+ ({k}\cdot{q_2} )^2 )-\Delta ^2 Q^2 {g}^{\mu \nu }-4
          {k'}^{\mu }  {k'}^{\nu }(    {k}\cdot  {q_2} )+4        {k}^{\mu }       {k}^{\nu }
        (  {k'}\cdot {q_2} )+2 (  {k}\cdot {\Delta } ) (   {k}^{\nu }{q_2}^{\mu }+   {k}^{\mu }   {q_2}^{\nu } ) \bigg\} \, .
	\end{align}
	Contracting the above two tensors, one gets
	\begin{equation}
	    \mathcal{T}_{BH}^2 = \frac{e^6}{(1+\epsilon^2)^2 x_B^2 y^2\Delta^2 P_1(\phi) P_2(\phi)} \bigg\{  c_0(\bar K) +c_1(\bar K) \cos(\phi)  + c_2(\bar K)\cos(2 \phi ) \bigg\} \,. 
	\end{equation}
	where $\bar K =\{x_B, \Delta^2,Q^2,M\}$ accounts for the dependence of the coefficients $c_i$ upon the kinematical invariants of the process, 
explicitely given, e.g., in Ref.
\cite{Belitsky:2001ns}.

	\subsubsection{Interference term}
	Since it is linear in the CFFs and allows the experimental extraction of these functions,
the interference term 
	 \begin{equation}\label{int}
	 \mathcal{I}_{BH-DVCS} =2 \Re e [\mathcal{T}_{DVCS}\mathcal{T}_{BH}^*]
	 \end{equation}
is the most interesting quantity
for GPDs phenomenology.
	 	The interference amplitude, in terms of leptonic and hadronic tensors, reads
	\begin{eqnarray}\label{int1}
	\mathcal{I}_{BH-DVCS}= \frac{e^6}{\Delta^2 q_1^2}(-g_{\mu \mu'}) \sum_{S's'} (L^{DVCS}_\nu L^{BH}_{\mu' \rho} T^{\mu \nu} {\cal J}^{\rho \dagger} + \text{c.c} )
	=
		 \frac{e^6}{\Delta^2 q_1^2}(-g_{\mu \mu'}) \sum_{S'} ({\cal L}_{\nu \mu' \rho} T^{\mu \nu} {\cal J}^{\rho \dagger} + \text{c.c} )\, .
    \end{eqnarray}	
    
    The amplitude of the pure DVCS 
	process, $\mathcal{T}_{DVCS}$, depicted in Fig. \ref{dvcsinco}, is related to the DVCS hadronic tensor $T_{\mu \nu}$ given by the time-ordered product of the electromagnetic currents  $j_\mu(z) = e \sum_q \epsilon_q \bar{\psi}_q(z) \gamma^\mu \psi_q(z)$ of quarks with a fractional charge ($\epsilon_q$) sandwiched between hadronic states with different momenta (see, for details, 
	Ref. \cite{Belitsky:2001ns}).
The most general expression	for the hadronic tensor
$T_{\mu \nu}$, which
can be decomposed in a complete basis of CCFs $\mathcal{F}$ that, up to twist three, reads
	\begin{equation}
	\mathcal{F}(\xi,\Delta^2,Q^2) = \{ \mathcal{H },\mathcal{E}, \mathcal{\tilde{H}},\mathcal{\tilde{E}},\mathcal{H}_+,\mathcal{E}_+,\mathcal{\tilde{H}}_+, \mathcal{\tilde{E}}_+ \}\, ,
	\label{cffstmunu}%
	\end{equation}
has been worked out in Ref.
\cite{Belitsky:2001ns} and, at leading twist,
for an unpolarized target, at JLab kinematics,
can be approximated as
	\begin{equation}\label{adrodvcs}
	T_{\mu \nu} \simeq - \mathcal{P}_{\mu \sigma} g_{\sigma \tau} \mathcal{P}_{\tau \nu } \frac{q \cdot V_1}{P \cdot q } \, ,
	\end{equation}

with the projector operator
\begin{equation}\label{key1}
\mathcal{P}_{\mu \nu} = g_{\mu \nu} - \frac{q_{1 \mu} q_{2 \nu }}{q1 \cdot q2}\, ,
\end{equation}
which ensures  current conservation, since $ q_\mu \mathcal{P}^{\mu \nu}  =0$, and 
\begin{eqnarray}
{V}_{ 1 \rho}= P_\rho \frac{ q\cdot h}{q \cdot P }\mathcal{H}(\xi, \Delta^2) + P _ \rho 
\frac{q \cdot e}{q \cdot P } \,  \mathcal{E} (\xi, \Delta^2) \, .
\label{v1rho}
\end{eqnarray}
The above expression is given in terms of CFFs and Dirac bilinears, defined as follows \cite{Belitsky:2001ns}
 \begin{eqnarray}\label{key2}
h_\rho &=& \bar{u}(p_N,S')\gamma_\rho u(p_1,S) \, , \\
e_\rho &=& \bar{u}(p_N,S')  i \sigma_{\rho \nu} \frac{\Delta_\nu}{2 M} u(p_1,S)
\, .
\end{eqnarray}
Using \eqref{adrodvcs}
- \eqref{v1rho}
, a term appearing
in Eq. \eqref{int1}, after summation over the final proton polarizations, can be effectively cast 
in the following way
\begin{equation}\label{adroint}
\sum_{S'}\frac{q \cdot V}{P\cdot q} {\cal J}^{\rho \dagger}  + \text{c.c} = P_\rho \,    [ \mathcal{C}^{int}_{unp} 
( \mathcal{F})] + 2 q_\rho \frac{\Delta^2}{Q^2} \mathcal{C}^{int, vec}_{unp}( \mathcal{F}) \, ,
\end{equation}
where we introduced the following combination of CFFs
\begin{eqnarray}\label{cff}
{C}^{int}_{unp}( \mathcal{F}) &=& F_1 \mathcal{H}(\xi,\Delta^2) -\frac{\Delta^2}{4 M ^2} F_2 \mathcal{E}(\xi,\Delta^2) \, , \\
 \mathcal{C}^{int, vec}_{unp} ( \mathcal{F})&=& \xi (F_1 + F_2) ( \mathcal{H}(\xi,\Delta^2)+ \mathcal{E}(\xi, \Delta^2)) \, .
\end{eqnarray}
As everywhere in this paper, the dependence of the CFFs on the 
scale $Q^2$ is omitted.
{ After contracting the leptonic and the hadronic tensors, the interference term can be decomposed in harmonics, i.e.}
\begin{equation}
    \mathcal{I}_{BH-DVCS} = \frac{e^6}{y^3 x_B\Delta^2P_1(\phi)P_2(\phi)} (c_0^\mathcal{I} +\sum_{n=1}^3 c_n^\mathcal{I} \cos( \phi) +s_n^\mathcal{I} \sin(n\phi)) \, .
\end{equation}
{As it can be read in the expressions explicitely given in Ref. \cite{Belitsky:2001ns},
the only terms not suppressed 
at JLab kinematics
are $c_1^{\cal I}$ and  $s_1^{\cal I}$,
with the latter clearly dominating the former.
Besides, in the BSA, only $s_1^{\cal I}$, linear
in $\lambda$, appears.
We therefore consider it as the only relevant
contribution to the interference.
} In particular, it turns out that $s_1^{\cal I}$ depends only on the combination of CFFs given in \eqref{cff}, with the term
proportinal to $\cal H$ clearly dominating at JLab kinematics.
Therefore in the following we consider $\cal H$ as the only relevant CFF.
For later convenience, we notice
that the only part of the leptonic tensor in Eq. \eqref{int1} which is ontributing
to the $s_1^{\cal I}$ term is 
  \begin{eqnarray}
  \bar {\mathcal{L}}^{\mu \nu \rho} &= -2 i \lambda Q^2 (2 P_1(\phi) g ^{\mu \nu } \epsilon ^{\rho k k' q_2}-2 P_2(\phi) g^{\nu \rho } \epsilon ^{\mu k k' q_2}-2 P_1(\phi) k^{\rho } \epsilon ^{\mu \nu k' q_2}+2 P_1(\phi) q_2^{\rho} \epsilon ^{\mu \nu k k'}-2 P_1(\phi) k'^{\mu } \epsilon ^{\nu \rho k q_2} - 
  \nonumber \\ 
  \nonumber
  &2 P_1(\phi) k'^{\nu } \epsilon ^{\mu \rho k q_2}-2 P_1(\phi) ( k\cdot q_2 ) \epsilon ^{\mu \nu \rho k'}+2 P_1(\phi) k'^{\rho } \epsilon ^{\mu \nu k k'}-2 P_1(\phi) k'^{\mu } \epsilon ^{\nu \rho k k'}-2 P_1(\phi) k'^{\nu } \epsilon ^{\rho \mu  k k'}-\\
  \nonumber
  &2 P_2(\phi) k'^{\nu } \epsilon ^{\mu \rho k q_2}-2 P_2(\phi) k'^{\rho } \epsilon ^{\mu \nu k q_2}-2 P_2(\phi) k^{\mu } \epsilon ^{\nu \rho k' q_2}+2 P_2(\phi) q_2^{\mu } \epsilon ^{\nu \rho k k'}-2 P_2(\phi) ( k\cdot q_2)\epsilon ^{\mu \nu \rho k'}- \\
  &4 P_2(\phi) k^{\mu} \epsilon ^{\nu \rho k k'}-P_1(\phi) Q^2 \epsilon^{\mu \nu \rho k'}) .
  \label{lepex}
  \end{eqnarray} 

Explicitely, one gets
$
s_1^{\cal I}=
8 \lambda \tilde K  y(2-y) \Im m 
( F_1(\Delta^2) \mathcal{H}(\xi,\Delta^2) ) 
$ 
and therefore

\begin{equation}\label{sint}
{\mathcal{I}_{BH-DVCS}} = \frac{ 8 \lambda e^6 \tilde K  (2-y) \sin \phi}{y^2 x_B P_1(\phi) P_2(\phi) \Delta^2} \Im m \bigg ( F_1(\Delta^2) \mathcal{H}(\xi,\Delta^2) \bigg) \, .
\end{equation}

If one considers
corrections of order $\epsilon^2$
and $\Delta^2/Q^2$, both coming from the leptonic part, it reads 
{
\begin{equation}
    \mathcal{I}_{BH-DVCS}= \frac{8 \lambda e^6 \tilde K \sin \phi}{P_1(\phi) P_2(\phi)\Delta^2 x_B y^2(1+\epsilon^2)^{\frac{3}{2}} } (2 J +4 \tilde K \cos \phi +y(1+\epsilon^2)) \Im m \bigg ( F_1(\Delta^2) \mathcal{H}(\xi,\Delta^2) \bigg) \, .
\end{equation}
{We used this formula for the interference part in the present calculation in order to have a coherent comparison between results for the bound proton and for the free one.}

	\subsection{Generalization to Deeply Virtual Compton Scattering off a moving off-shell proton }

First of all, let us define the components 
of the bound off-shell proton

\begin{align}\label{kinmoto}
p&= (p_0,|\vec{p}| \sin \vartheta \cos \varphi, |\vec{p}|\sin \vartheta \sin \varphi, |\vec{p}|\cos \vartheta) 
\end{align}
where $p_0 \neq \sqrt{M^2+|\vec p|^2} $ (see Eq. \eqref{off}).

	\subsubsection{Bethe Heitler term}
	Our goal is to obtain a formula for the BH contribution which generalizes the harmonic decomposition obtained for a proton at rest, well known in the literature.
	So, first, let us consider the general expression for Bethe Heitler amplitude given by Eq. \ref{tbh0}.
	In the square of the above mentioned amplitude,
	after summation over the final proton polarizations,
	the hadronic part reads
	
		\begin{align}
	 \sum_{S'} {\cal J}^\mu {\cal J}^{\dagger \nu}& = \frac{1}{2} \bigg [F_1(\Delta^2)^2+(F_1(\Delta^2)+F_2(\Delta^2))^2-\frac{\Delta ^2 }{4 M^2} F_2(\Delta^2)\bigg] (p^{\nu }
   p_N^{\mu }+p^{\mu } p_N^{\nu })+\frac{\Delta ^2}{2}  (F_1(\Delta^2)+F_2(\Delta^2))^2 g^{\mu \nu }+ \nonumber \\ 
   & \frac{1}{2}
   \bigg(F_1(\Delta^2)^2-(F_1(\Delta^2)+F_2(\Delta^2))^2-\frac{\Delta ^2 }{4 M^2}F_2(\Delta^2)\bigg) (p^{\mu } p^{\nu
   }+p_N^{\mu } p_N^{\nu }) + (-\Delta^2 + 2 M^2 - 2 p \cdot p_N) \bigg[\frac{g^{\mu \nu }}{2} *  \nonumber \\
    &  \bigg (F_1^2(\Delta^2) -\frac{F_2(\Delta^2)^2}{2} +  F_2(\Delta^2)\frac{ p\cdot p_N }{2 M^2} \bigg) - \frac{F_2(\Delta^2)^2}{8 M^2} \bigg((p_N^\mu p^\nu +p_N^\nu p^\mu) - (3 p_N^\mu p_N^\nu +p^\mu p^\nu) \bigg ) \bigg] \, .
	\end{align}
This expression accounts for the motion of the initial proton and reduces to the one obtained for a  proton at rest given by Eq. \eqref{adrorest} when $p_0 \rightarrow M, \vec p \rightarrow \vec 0$. 

	As for the lepton propagators, we have the same structure of Eqs. \eqref{propbh}, i.e.
		\begin{eqnarray}
	    P_1(\phi) & = & 1+\frac{2(\mathcal{J}({K_b})-\mathcal{K}({K_b})\cos \phi) }{Q^2} \, ,  \nonumber \\ 
	    P_2(\phi) & = & \frac{\Delta^2 - 2(\mathcal{J}( K_b)-\mathcal K(K_b)\cos\phi) }{Q^2} \, ,
	    \label{propabhmoto}
	\end{eqnarray}
	but $\mathcal{J}$ and $\mathcal{K}$ become functions of the invariant kinematical variables and of the 4-momentum components of the initially moving bound proton, i.e ${K_b} = \{M, x_B, \Delta^2,Q^2,\vec p ,p_0\}$:
	\begin{align}\label{jkmoto}
	    \mathcal{J}(K_b)&= E_k(E_2-p_0 - \cos \theta_e(|\vec{p}_N| \cos\theta_N - |\vec{p} |\cos\vartheta ) +  |\vec{p}|\sin \theta_e \sin \vartheta \cos\varphi )  \\
	    \mathcal{K}({K}_b)&= E_k \sin \theta_e |\vec{p}_N| \sin\theta_N 
	    \, .
	\end{align}

	 With these ingredients at hand, one can compute the full contraction between the leptonic contribution \eqref{leptobh} and the hadronic one for the BH process. In this way, 
	 a long and complicated analytical expression is obtained \cite{saratesi}. It is not reported here
	 but the interested readers can obtain either a Mathematica notebook or a Fortran code from the authors upon request.
	 The scalar products there appearing have to be evaluated considering the motion of the initial nucleon and its off-shellness. 	If one evaluates instead the scalar products for a proton at rest, the obtained expression reduces to the one of the previous section for a proton at rest, as expected.
	
	\subsubsection{Interference term}

    The BH-DVCS interference term for a
    moving proton will be given, as always, by the
    contraction of a lepton and a hadronic tensor.
	The leptonic part is the same already obtained for a proton at rest and written in Eq. \eqref{lepex}, but now the lepton propagators have to evaluated according to Eq.
	\eqref{propabhmoto}.
	
	Concerning the hadronic tensor, we obtain the following result 
	for the contribution
	Eq.\eqref{adroint} when the off-sehell proton
	is moving
	\begin{equation}\label{adroexmoto}
	\sum_{S_2}\frac{q \cdot V}{P\cdot q} J^{\rho \dagger}  + \text{c.c} = P_\rho \,\mathcal C^{int}_{unp} ( \mathcal{F}) + 2 q_\rho \frac{\Delta^2}{Q^2} \mathcal C^{int \, ,vec}_{ unp}(\mathcal{F})
	\end{equation}
	where the the combination of CFFs has to be read: 
	\begin{eqnarray}\label{key3}
	\mathcal C^{int}_{unp}(\mathcal{F}) &= F_1 (\Delta^2)\mathcal{H} (\xi ,\Delta^2)- F_2(\Delta^2) \mathcal{E} (\xi ,\Delta^2)\frac{\Delta^2}{4 M^2}\bigg[ 1+ {\xi}\bigg(\frac{\Delta^2- 4 M^2 + 2  p\cdot p_N}{\Delta^2} \bigg) \bigg]  \, , \\
		\mathcal C^{int \, ,vec}_{unp} (\mathcal{F}) &=\xi \bigg[ F_1 (\Delta^2)\mathcal{H}(\xi, \Delta^2) \bigg( 1 +\frac{M^2 -|\vec p|^2}{\Delta^2}\bigg) +
		F_1 (\Delta^2) \mathcal{E} (\xi ,\Delta^2)+F_2 (\Delta^2)\mathcal{H} (\xi, \Delta^2)+ \nonumber \\
		& F_2(\Delta^2) \mathcal{E} (\xi, \Delta^2)  \bigg(\frac{3}{2} +  \frac{ p\cdot  p_N}{2 M^2} - \frac{M^2}{\Delta^2} + 2 \frac{p \cdot p_N}{\Delta^2} - \frac{p \cdot p_N}{M^2} (1+ \frac{p \cdot p_N}{\Delta^2})\bigg) \bigg] \, , \nonumber
	\end{eqnarray}
	where use has been made of $ \Delta \cdot q \approx - \xi (P \cdot q )$ and, for the relevant scalar product, one has 
	$p \cdot p_N = p_0 E_2 - |\vec{p}| \, |\vec{p}_N|\cos(\theta_{\hat{p p_N}}) $.
	In order to get the explicit expression for the only term  appearing in the 
	interference, the contraction between the leptonic part, given by Eq. \eqref{lepex}, and the hadronic tensor, Eq. \eqref{adroexmoto}, has to be performed.
	Also here,
in the actual calculation we are considering the dominance  of $\mathcal{H}(\xi,\Delta^2)$.
		The final result reads:
\begin{multline}
{\cal I}_{BH-DVCS} = \frac{4 \lambda \sin(\phi)}{Q^4 \Delta^2 P_1(\phi)P_2(\phi) y \epsilon^2}\bigg(3 ( P_2(\phi) - P_1(\phi)) + P_2(\phi) ^2+P_1(\phi)^2 \bigg )  \\ 
 \bigg(2 | \vec p_N | Q^2 \sin\theta_N (p_0 \sin\theta_e  \sqrt{1+\epsilon^2} + | \vec p|  \sin \theta_e \cos \vartheta - |\vec p| (\cos \theta_e + \sqrt{1+ \epsilon^2}) \cos\varphi \sin\vartheta)\,\bigg) \Im m \bigg[F_1(\Delta^2) \mathcal{H}(\xi',\Delta^2) \bigg]\, ,
\end{multline}
	where the propagators $P_{1,2}(\phi)$ are again given by Eqs. \eqref{propbh} with the proper definition of the quantities appearing in there and given by Eqs. \eqref{propabhmoto}.
	Nuclear effects on the parton content of the bound proton appears only in the CFF, which has to be evaluated properly using the skewness $\xi' = \left [ {Q^2(1+\frac{\Delta^2}{2 Q^2})} \right ]/(2 {P \cdot q})$, accounting for the motion of the bound proton in the nuclear medium. 	
   
   Therefore, using the above interference term
   and the one discussed in the previous subsection for the squared of the BH amplitude,
   we can evaluate the cross sections
   \eqref{crossl},
   for a given kinematic and electron helicity and, in turn, the beam spin asymmetries and all the results
   shown in this paper.
	\newpage

\twocolumngrid

\end{document}